\documentclass[preprint,showpacs,preprintnumbers,amsmath,amssymb,endfloats]{revtex4}

\usepackage{graphicx}
\usepackage{dcolumn}
\usepackage{bm,subfigure}

\begin{document}

\preprint{APS/123-QED}

\title{Comparison of Langevin and Markov channel noise models for neuronal signal generation }

\author{B. Sengupta}
\email{bs393@cam.ac.uk}
\author{S. B. Laughlin}
\author{J. E. Niven} 
\email{jen22@hermes.cam.ac.uk}
\altaffiliation{Smithsonian Tropical Research Institute, Apartado 0843-03092, Balboa, Anc\'{o}n, Panam\'{a}, República de Panam\'{a}} 
\affiliation{%
Neural Circuit Design Group, Department of Zoology, University of Cambridge\\
Downing Street, Cambridge CB2 3EJ, United Kingdom
}%

\date{\today}

\begin{abstract}
Abstract --- The stochastic opening and closing of voltage-gated ion channels produces noise in neurons. The effect of this noise on the neuronal performance has been modelled using either approximate or Langevin model, based on stochastic differential equations or an exact model, based on a Markov process model of channel gating. Yet whether the Langevin model accurately reproduces the channel noise produced by the Markov model remains unclear. Here we present a comparison between Langevin and Markov models of channel noise in neurons using single compartment Hodgkin-Huxley models containing either $Na^{+}$ and $K^{+}$, or only  $K^{+}$ voltage-gated ion channels. The performance of the Langevin and Markov models was quantified over a range of stimulus statistics, membrane areas and channel numbers. We find that in comparison to the Markov model, the Langevin model underestimates the noise contributed by voltage-gated ion channels, overestimating information rates for both spiking and non-spiking membranes. Even with increasing numbers of channels the difference between the two models persists. This suggests that the Langevin model may not be suitable for accurately simulating channel noise in neurons, even in simulations with large numbers of ion channels.
\end{abstract}

\pacs{87.10.Mn,87.19.lo,87.19.lc}
\maketitle

\section{\label{sec:intro}Introduction}

Signalling in neural circuits is constrained by the presence of noise. Extrinsic noise, such as photon shot noise, is present in sensory inputs but noise is also generated intrinsically by molecular components within neurons \cite{Faisal2008}. One major source of intrinsic noise, channel noise, is a consequence of the stochastic opening and closing of voltage-gated ion channels found in neural membranes \cite{Neher1976,Lecar1971a,Lecar1971b}. These voltage-gated ion channels are themselves responsible for both filtering and generating electrical signals within neurons. Stochasticity in voltage-gated ion channels is caused by random fluctuations between different conformational states due to thermal agitation \cite{White2000}. The effect of stochasticity becomes increasingly important in neurons with small numbers of ion channels because channel noise declines in proportion to the square root of the number of channels  in a membrane \cite{White2000}. In neurons capable of supporting action potentials (spikes), channel noise can alter the reliability of single spikes, their spiking threshold and their firing rate \cite{Schneidman1998,Clay1983,Schmid2004,Skaugen1979}. It can also cause neurons to display stochastic resonance by amplifying external signals \cite{Schmid2001}.

Channel noise can be modelled either using  exact  \cite{Strassberg1993,Rubinstein1995,Chow1996} or approximate \cite{Fox1994,Fox1997} methods for the transformation of ion channel state changes (open versus closed) into fluctuations in membrane potential. Exact or Markov methods model channel noise as continuous time Markov processes to iterate through the transition probability matrix of state change to infer the exact number of ion channels opened during each time step \cite{Fall2002}. Approximate or Langevin methods model channel noise as an additive Gaussian noise perturbation of the activation and inactivation variables of the voltage-gated ion channels \cite{Fall2002}. Exact methods may be computationally demanding, slowing simulations and making approximate methods more favourable  \cite{Mino2002}. However, it is unclear the extent to which approximate methods capture the stochastic fluctuations of voltage-gated ion channels and their impact on neural signalling. Comparison with simulations using exact channel noise models suggests that the approximate method fails to predict the number, latency and jitter of action potentials generated in response to steps or pulses of current injection  \cite{Mino2002,Bruce2007,Bruce2009}. However, experiments show that the spike trains generated by repeated presentations of a fluctuating input current are more reproducible between trials than those generated by a constant input current \cite{Mainen1995}. Thus, knowing the differences between channel noise models in response to steps or pulses of current is not sufficient to predict the differences between exact and approximate channel noise models in response to fluctuating inputs. 

We use single compartment Hodgkin-Huxley models in which either the exact or approximate models were implemented to simulate the stochastic opening and closing of voltage-gated ion channels. These models were stimulated with presentations of fluctuating input currents allowing their performance and reliability to be quantified in terms of their information rates \cite{Rieke1997}. The information coded was quantified in single compartment models that support action potentials (spiking models) as well as those supporting only analogue signals (non-spiking models) because both types of neurons are found in vertebrate and invertebrate nervous systems \cite{Roberts1981}. We varied the size of the single compartment being simulated to determine whether differences between the exact and approximate methods are greater in smaller compartments with fewer voltage-gated ion channels where the effects of channel noise are more pronounced \cite{Rubinstein1995,Faisal2005}. We also compared the performance of the exact and approximate methods in the presence or absence of extrinsic noise. We show that single compartments with the Langevin (approximate) channel noise model have higher information rates than compartments of identical size with the Markov model (exact) of channel noise. Furthermore, we show that this overestimation of information rate by the approximate method is due to underestimation of the power of the intrinsic noise that does not improve even in larger compartments with greater numbers of ion channels. 

\section{\label{sec:model}Methods}

\subsection{\label{sec:singcmpt} Single Compartment Model}

We used a single compartment stochastic Hodgkin-Huxley model for our simulations \cite{Skaugen1979}. The spiking model contained two voltage-gated ion channels, $Na^{+}$ and delayed rectifier $K^{+}$ along with the leak conductance while the non-spiking model only possessed delayed rectifier $K^{+}$ and  leak conductances. The dynamics of the membrane potential was governed by a set of activation and inactivation variables, $m_j$ and $h_j$ with the current balance equation that had  the general form,
\begin{eqnarray}
{C_m}\frac{{dV_m}}
{{dt}} = \sum\limits_j {{g_j}} m_j^{{a_j}}h_j^{{b_j}}\left( {{E_j} - {V_m}} \right) + {I_{stim}(t)} + {\zeta _{noise}(t)},
\end{eqnarray}

where $C_m$ was the membrane capacitance, $g_j$ was the conductance of the $j^{th}$ conductance type, $a_j$ and $b_j$ were integers, $E_j$ was the reversal potential of the $j^{th}$ conductance, $I_{stim}(t)$ was a time dependent current stimulus and $\zeta_{noise}(t)$ was the extrinsic stimulus noise current. $\zeta_{noise}(t)$ was zero for no input noise simulations. The variables $m_j$ and $h_j$ followed first order kinetics of the form $\frac{{dm}}
{{dt}} = \frac{{{m_\infty }({V_m}) - m}} {{\tau (V_m)}}$, where $m_\infty ({V_m})$ was the steady-state activation and $\tau (V_m)$ was the voltage-dependent time constant. The model was driven using a time dependent current - $I_{stim}(t)$, which was either a 500 Hz $(\tau_{correlation} = 2 \ ms)$ or  a 300 Hz $(\tau_{correlation} = 3.3 \ ms)$ Gaussian white noise, filtered using a $40^{th}$-order Butterworth filter to approximate a box filter in the frequency domain. The mean and the standard deviation of the stimulus was varied in the range 1-10 $\mu A / cm^2$. The stimulus was presented for 1 second and each set of simulations consisted of 60 such trials. $\zeta_{noise}(t)$ was an unfiltered broad-band Gaussian white noise with,

\begin{eqnarray}
\left\langle {\zeta_{noise} (t)} \right\rangle  & = & 0  \nonumber  \\  
\left\langle {{\zeta _{noise}}(t){\zeta _{noise}}(t')} \right\rangle  & = & {\sigma ^2}\delta (t - t'),
\end{eqnarray}

where, noise variance was computed using

\begin{eqnarray}
\sqrt {\frac{{\overbrace {\frac{1}
{T}\int\limits_0^T {I_{stim}(t)^2} dt}^{Signa{l_{RMS}}}}}{SNR}} .
\end{eqnarray}

All Gaussian random numbers were generated using the Marsaglia's ziggurat algorithm \cite{Marsaglia2000}; uniform random numbers were generated using Mersenne Twister algorithm \cite{Matsumoto1998}. Deterministic equations were integrated using the Euler-algorithm while stochastic differential equations were integrated using the Euler-Maruyama method \cite{Kloeden2000}, both with a step size of 10 $\mu s$.  Parameter values are given in Table \ref{table:nonlin}.

\begin{table*}[ht] \caption{Parameters for the stochastic Hodgkin-Huxley model \cite{HODGKIN1952}.} 
\centering   
\begin{tabular}{|l|l|l|}
\hline
Symbol & Definition & Value, units \\
\hline
\hline
$C_{m}$ & Specific membrane capacitance & $1 \ \mu F /cm^2$ \\ 
T &  Temperature & $6.3\,^{\circ}\mathrm{C}$ \\ 
$E_{l}$ & Leakage reversal potential & $-54.4 \ mV$ \\ 
$E_{Na}$ & Sodium reversal potential & $50 \ mV$ \\ 
$E_{K}$ & Potassium reversal potential & $-77 \ mV$ \\ 
$g_{l}$ & Leakage conductance & $0.3 \ mS/cm^2$ \\
$g_{Na}$ & Sodium channel conductance & $20 \ pS$ \\
$g_{K}$ & Potassium channel conductance & $20 \ pS$ \\
$N_{Na}$ & Sodium channel density & $60\ / \ \mu m^2$ \\
$N_{K}$ & Potassium channel density & $18 \ / \ \mu  m^2$ \\
A & Area of the cell & $1,10,100 \ \mu m^2$\\
$\alpha_{m}(V)$ &  opening rate (activation, $Na^{+}$) & $\frac{{0.1({V_m} + 40)}} {{1 - \exp ( - 0.1({V_m} + 40))}}$ \\
$\alpha_{h}(V)$ &  opening rate (inactivation, $Na^{+}$)& $0.07\exp ( - 0.05({V_m} + 65))$ \\
$\alpha_{n}(V)$ &  opening rate (activation, $K^{+}$) & $\frac{{0.01({V_m} + 55)}} {{1 - \exp ( - 0.1({V_m} + 55))}}$ \\
$\beta_{m}(V)$ &  closing rate (activation, $Na^{+}$) & $4\exp ( - 0.0556({V_m} + 65))$ \\
$\beta_{h}(V)$ &  closing rate (inactivation, $Na^{+}$)  & $\frac{1} {{1 + \exp ( - 0.1({V_m} + 35))}}$ \\
$\beta_{n}(V)$ &  closing rate (activation, $K^{+}$) & $0.125\exp ( - 0.0125({V_m} + 65))$ \\
\hline
\end{tabular} 
\label{table:nonlin}  
\end{table*} 

\subsection{\label{sec:methods}Model of Channel Noise}

\subsubsection{Exact method}

\begin{figure*}[htbp]
\centering
\subfigure{\includegraphics[scale=0.6]{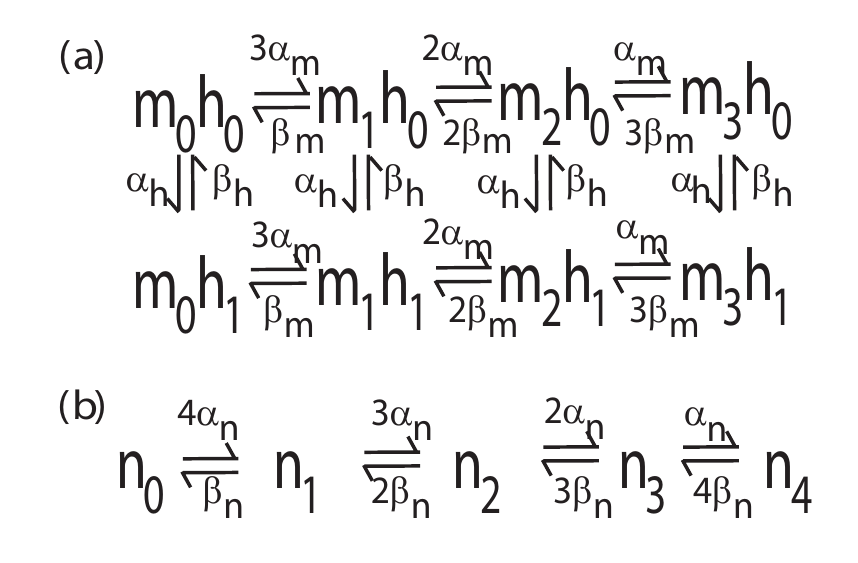}}
\caption{Markov state transitions for the voltage-gated ion channels. (a) Gating scheme for the $Na^{+}$ channel. (b) Gating scheme for the $K^{+}$ channel.}
\label{fig:NaK_gating}
\end{figure*}

We assumed that the $Na^{+}$ and the $K^{+}$ voltage-gated ion channels were not cooperative and that they had transitions between the closed and open states according to a Markov process (Fig. \ref{fig:NaK_gating}) \cite{Chow1996}. The number of voltage-gated ion channels in either the closed or the open state was tracked \cite{Gillespie1977,Chow1996}. At any time, the voltage-gated ion channels were distributed over 13 states with 28 possible transitions between these states - 20 transitions for the  $Na^{+}$ and 8 transitions for $K^{+}$ voltage-gated ion channels. When the voltage-gated ion channel was in state $k$ at time $t$, the probability that it would remain in that  state in time interval $\delta t$ was ${e^{ - {\gamma _i}\delta t}}$, where $\gamma _i$ was the sum of all transition rates from state $k$ to any possible successive state. During $\delta t$ no other voltage-gated ion channel changed its state. The probability of the ion channels remaining in the same state in the time interval $\delta t$ was ${e^{ - {\lambda}\delta t}}$, 

\begin{eqnarray}
\lambda  = \sum\limits_{i = 0}^3 {\sum\limits_{j = 0}^1 {\left[ {{m_i}{h_j}} \right]} } {\gamma _{ij}} + \sum\limits_{k = 0}^4 {\left[ {{n_k}} \right]} {\gamma _k}, 
\label{eqn:lam}  
\end{eqnarray}

where, $\left[ {{m_i}{h_j}}\right]$ was the number of $Na^{+}$ voltage-gated ion channels in state $ {{m_i}{h_j}}$, $ \left[ {{n_k}} \right]$ was the number of $K^{+}$ voltage-gated ion channels in state $ {{n_k}}$, ${\gamma _{ij}}$ was the total transition rate from state $ {{m_i}{h_j}}$,  ${\gamma _{k}}$ was the total transition rate from state $ {{n_k}}$. The transition rate $t_{trans}$ for a particular  ion channel state was chosen by drawing a pseudo-random number $r_1$ from an uniform distribution, $[0,1]$ and defining $t_{trans} = ln(r_1^{ - 1})/ \lambda$. The Gillespie algorithm then selects which of the 28 possible transitions occured in the time interval $t_{trans}$  \cite{Gillespie1977,Chow1996}. The conditional probability of a particular transition $j$ that occured in the time interval $\delta t$ was given by

\begin{eqnarray}
\frac{{{a_j}\delta t}}{{\sum\nolimits_{i = 1}^{28} {{a_i}\delta t} }} = \frac{{{a_j}}}{{\sum\nolimits_{i = 1}^{28} {{a_i}} }},
\end{eqnarray}

where, $a_j$ was the product of transition rate associated with transition $j$ and the number of channels in the original state of that transition. The denominator was equal to $\lambda$ (Eqn. \ref{eqn:lam}). The specific transition rate was selected by drawing a random number $r_2$ from the uniform distribution $[0,1]$ with $\psi$ defined as

\begin{eqnarray}
\sum\limits_{i = 1}^{\psi  - 1} {{a_i}}  < {r_2} \le \sum\limits_{i = 1}^\psi  {{a_i}} .
\end{eqnarray}

The number of voltage-gated ion channels in each state was updated and the membrane potential was re-calculated. Similar algorithm was used for the channel noise in the non-spiking neuron. 

\subsubsection{Approximate method}

Approximate channel noise implementation followed the Langevin formulation \cite{Fox1994,Fox1997}, which was based on the theory of stochastic differential equations \cite{Higham.2001}. The Langevin approximation is a continuous stochastic description of the gating kinetics of the ion channel; the Master equation governing the conductance of the cell is approximated by the Fokker-Planck equation \cite{Gardiner1996}. The gating variables evolved according to the following noise perturbations,

\begin{eqnarray}
\frac{{d\kappa }}{{dx}} = {\alpha _\kappa }(1 - \kappa ) - {\beta _\kappa }\kappa  + {\xi _\kappa }(t),
\end{eqnarray}

where $\kappa$ was $m$ or $h$ for the $Na^{+}$ channel and was $n$ for the $K^{+}$ channel,  $\xi_{\kappa }(t)$ was a Gaussian, zero mean perturbation to the activation and in-activation variables with the following variance,

\begin{eqnarray}
\left\langle {{\xi _\kappa }(t){\xi _\kappa }(t')} \right\rangle  & = & \frac{2}{{N_{\max }^\kappa }}\frac{{{\alpha _\kappa }(t)(1 - \kappa (t)) + {\beta _\kappa }(t)\kappa (t)}}{2}\delta (t - t') \nonumber  \\  
& \approx  &  \frac{2}{{N_{\max }^\kappa }}\frac{{{\alpha _\kappa }(t){\beta _\kappa }(t)}}{{{\alpha _\kappa }(t) + {\beta _\kappa }(t)}}\delta (t - t').
\end{eqnarray}

Here, $N_{\max }$ denoted the maximum number of ion channels of a particular type, $\kappa$ was either \textit{m,h or n}. The values of \textit{m,h and n} were restricted such that they do not leave the interval $[0,1]$ i.e., have a reflecting boundary.

\subsection{\label{sec:info}Calculation of Information Rate}
\subsubsection{Spiking neuron models}

We used the direct method to measure the entropy of the responses \cite{Ruyter1997,Strong1997}. This method involved, comparisons among different spike trains without reference to the stimulus parameters, which provided a direct measure of the amount of information contained in the neural response  without assumptions of what and how the information was represented in the neuron. The spike train entropy sets the information capacity for the spike train to carry information. The noise entropy on the other hand, measured the variability of the spike train across trials. These quantities were dependent upon the temporal resolution with which the spikes were sampled, $\Delta t$ and the size of time window, $T$. We used a different stimulus current presented in each subsequent trial (unfrozen noise) to calculate the spike train entropy, while using presentations of the same stimulus current in each subsequent trial (frozen noise) to calculate the noise correlation. We divided the spike train to form K-letter words with $K=T/\Delta\tau$. We used the responses from the unfrozen noise session, to estimate the probability of occurrence of particular word, $P(W)$. We estimated the total entropy as,

\begin{eqnarray}
\label{eqn:p_w_unfro}
S_{total}  =  - \sum\limits_W {P(W)\log _2 P(W)} \ \ bits .
\end{eqnarray}

We estimated the probability distribution of each word at specified time durations, $t$ so as to obtain $P(W|t)$. Entropy estimates were then calculated from these distributions and the average of the distributions at all times were computed to yield the noise entropy as,

\begin{eqnarray}
\label{eqn:p_wt_fro}
S_{noise}  = \left\langle { - \sum\limits_W {P(W|t)\log _2 P(W|t)} } \right\rangle _t \ \ bits ,
\end{eqnarray}

where, $ \left\langle \\ \right\rangle$ indicated average over time. The information was then computed as,

\begin{eqnarray}
\label{eqn:dirinfo}
I = S_{total}  - S_{noise} .
\end{eqnarray}

The spike train entropy and the conditional noise entropy diverge in the limit of $\Delta\tau \to 0$, their difference converges to the true finite information rate in this limit \cite{Strong1998}. We  used bias correction methods such that the estimation of entropy was less prone to sampling errors \cite{Treves1995}. Using $\Delta t = 1 \ ms$, we varied the spike trains to form words of different lengths. Using these entropy estimates, we extrapolated to infinite word length from four most linear values of the curve of entropy against the inverse of word length.

\subsubsection{Non-spiking neuron models}
We used an upper-bound method to calculate the maximum information transferable by the non-spiking signals \cite{Rieke1997}. This was done by imposing an upper limit on the information transferred by computing the channel capacity \cite{Cover2006}. This method assumed that the neuronal response and the neuronal noise had independent Gaussian probability distributions in the frequency domain and the noise was additive in nature. We defined the stimulus \textbf{S}, as the mean neuronal response obtained from a frozen noise experiment. The noise in each trial was calculated by removing  the average response from the individual responses \textbf{$R_{i}$}. This separated the deterministic quality of the code from that of the noise. Due to Gaussian assumptions, it required enough data to estimate the mean and variance of the Gaussian probabilities. Since, a Gaussian distribution has the highest entropy for a given variance, the actual information might be lower than this bound. In our simulations, both the response and the noise had an approximately  Gaussian distribution. We obtained the mean response power spectrum and the noise power spectrum using the multi-taper spectral estimator and computed their ratio to be the signal-to-noise ratio (SNR) \cite{Bendat2000,Press2007}. This is then used to compute the information for a dynamic Gaussian channel as,

\begin{eqnarray}
\label{eqn:dyn_gauss_ch}
I(S,R) = \int\limits_0^k {\log _2 [1 + SNR(f)]df} .
\end{eqnarray}

For our simulations, the limits of the integral were taken from  2.3 Hz to either 300.6 Hz or 500.49 Hz. The integral was evaluated using trapezoidal rule \cite{Press2007}. 

\section{\label{sec:results}Results}

We compared Langevin and Markov formulations of channel noise using single compartment models, possessing $Na^{+}$ and  $K^{+}$ voltage-gated ion channels along with additional leak conductances. We simulated the responses of the models to low-pass filtered Gaussian stimuli with different means and variances. The area of the compartment models was either 1, 10 or 100 $\mu m^{2}$ and within these compartments the specific density of the voltage-gated $Na^{+}$ channels was $60 /  \mu m^2$ and the voltage-gated $K^{+}$ channels was $18  /  \mu  m^2$.  


\begin{figure*}[htbp]
\centering
\subfigure{\label{fig:digi_300Hz_1a}\includegraphics[width=1.8in,height=2in]{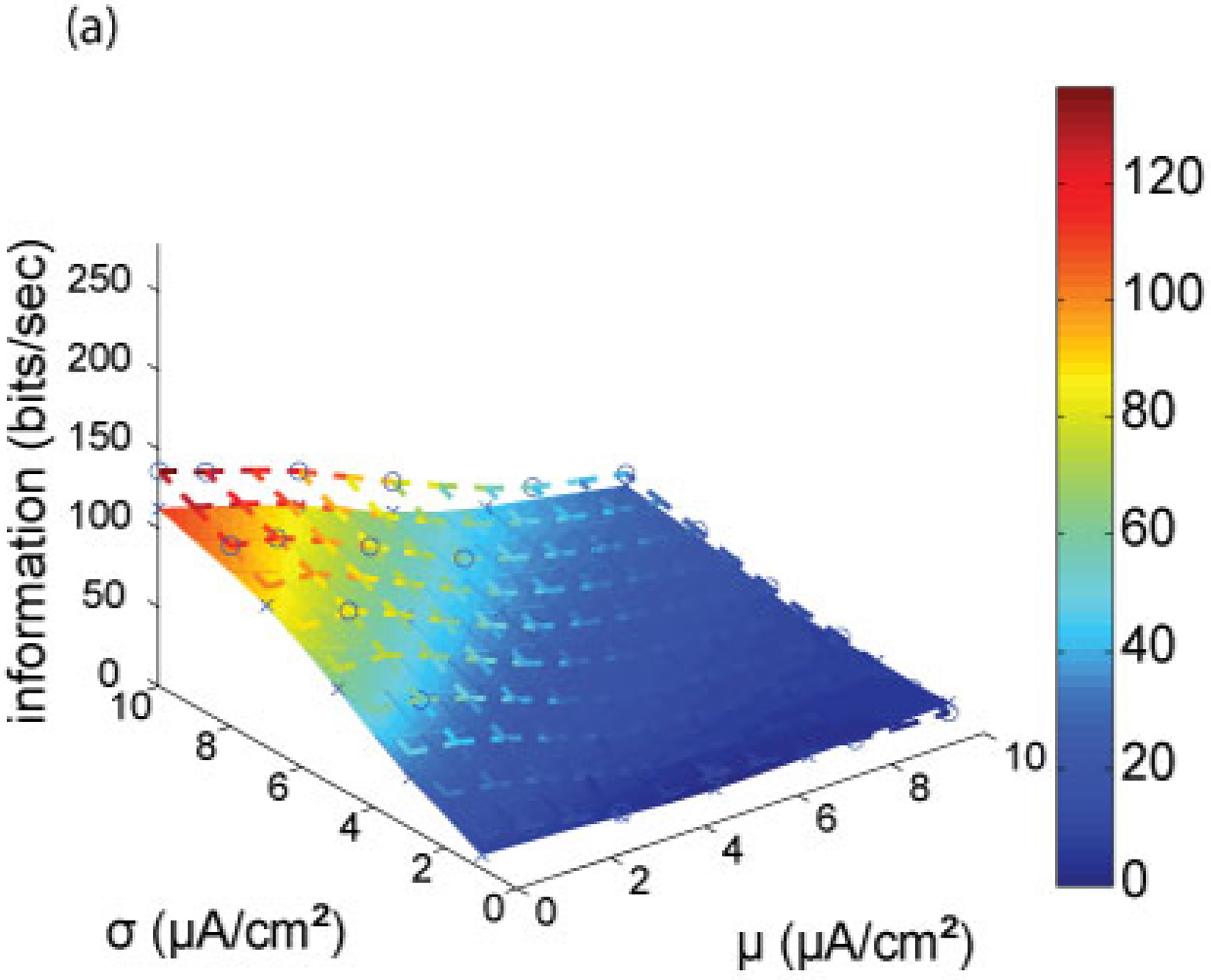}}
\subfigure{\label{fig:digi_300Hz_1b}\includegraphics[width=1.8in,height=2in]{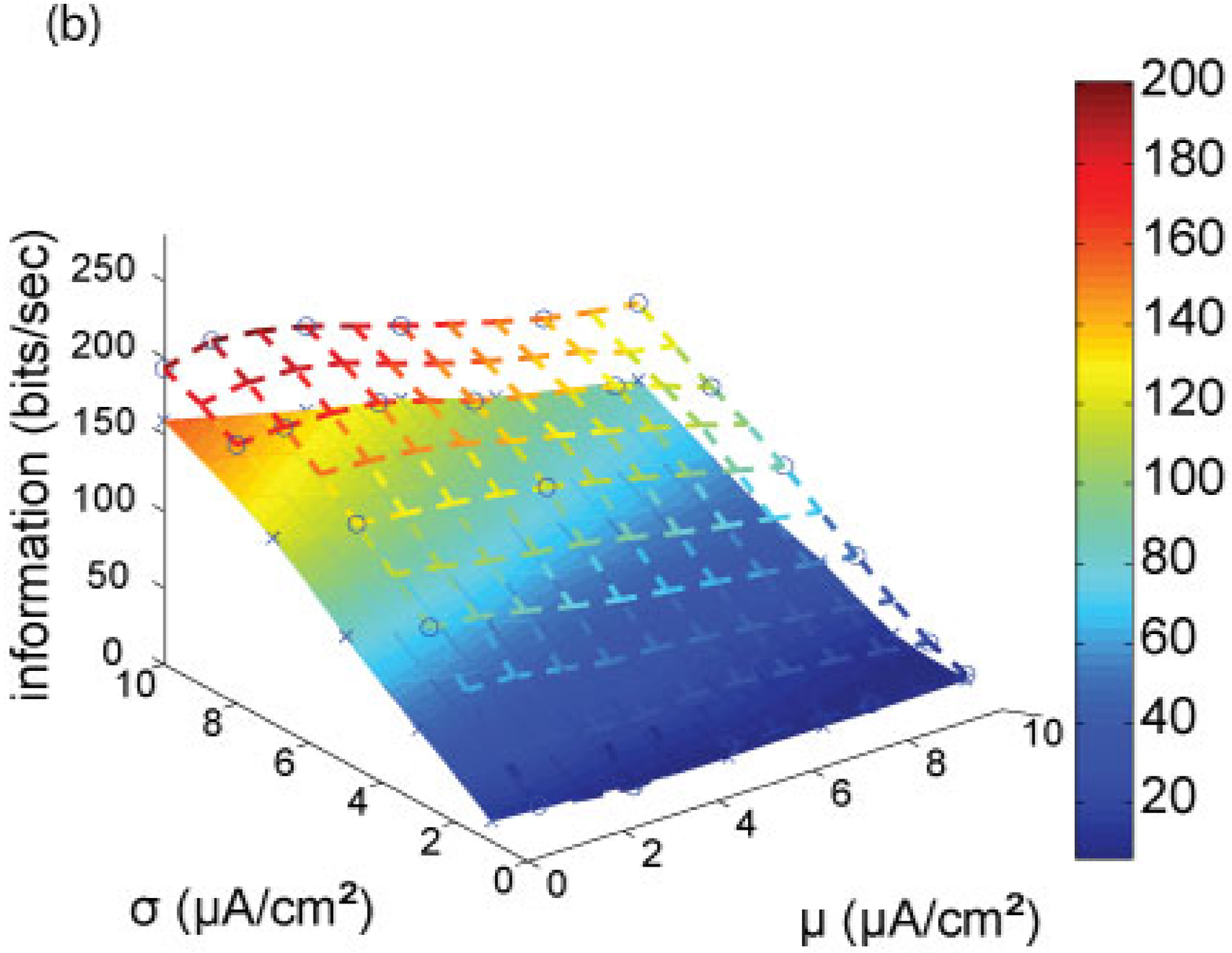}}
\subfigure{\label{fig:digi_300Hz_1c}\includegraphics[width=1.8in,height=2in]{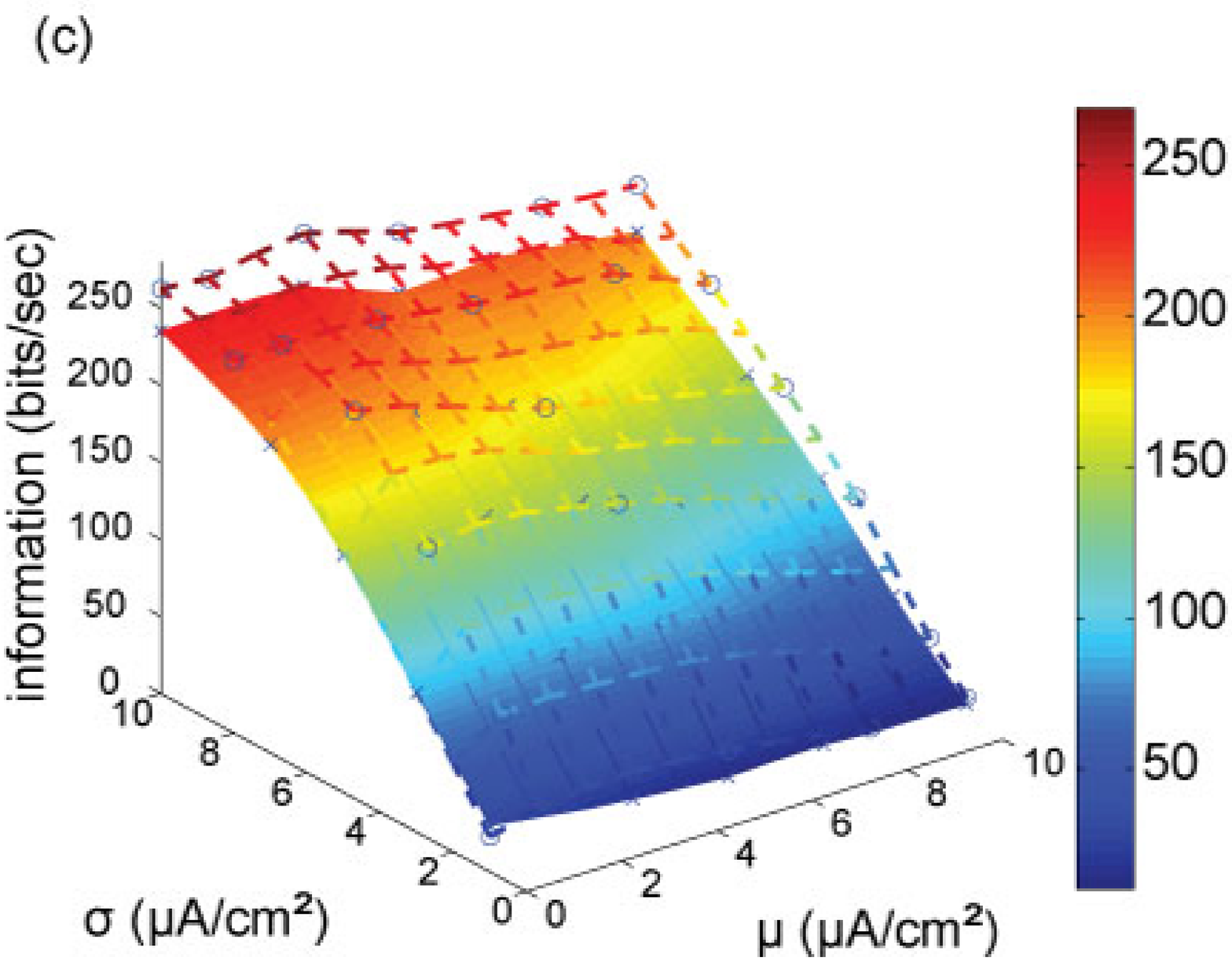}}\\
\subfigure{\label{fig:digi_300Hz_2a}\includegraphics[width=1.8in,height=2in]{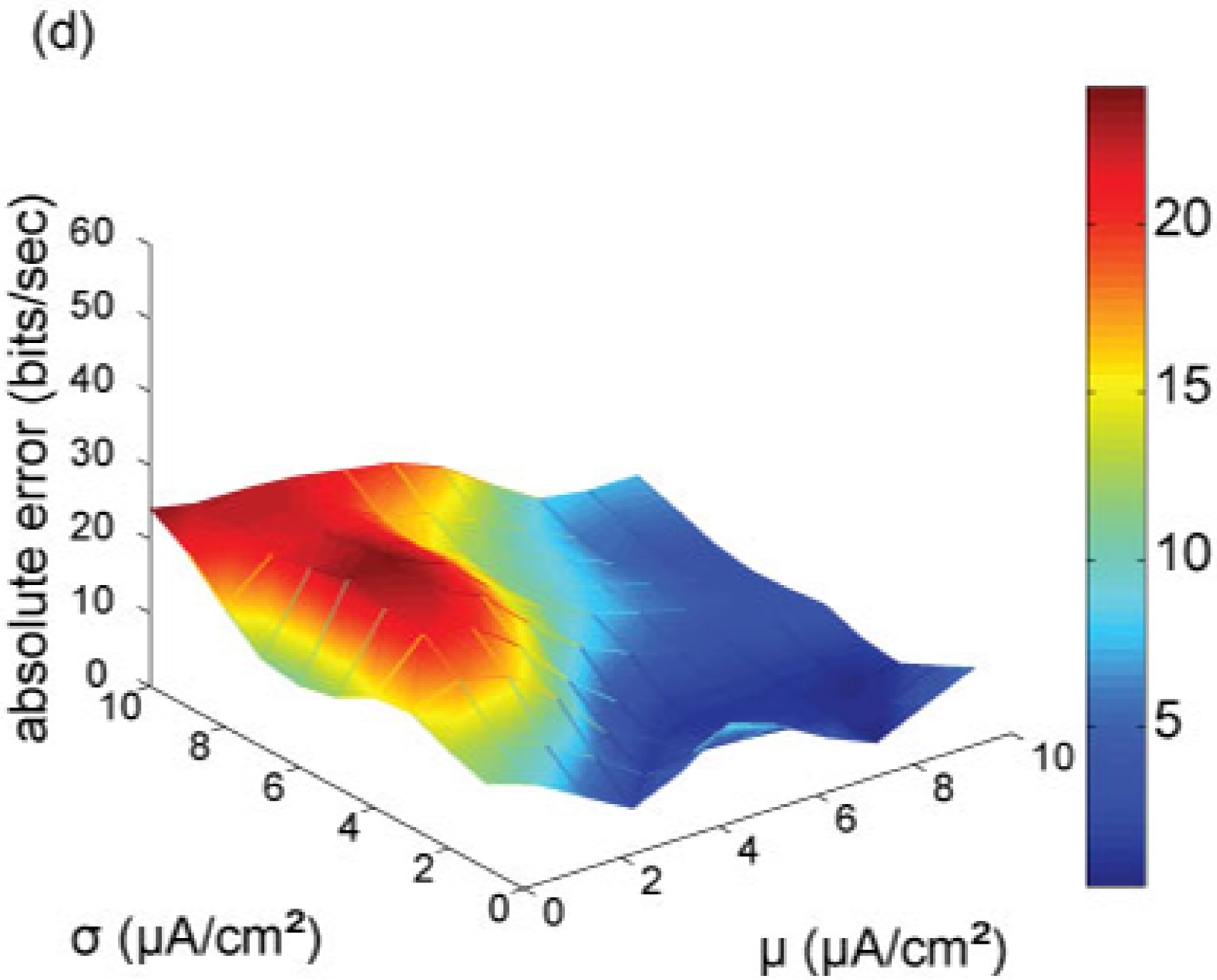}}
\subfigure{\label{fig:digi_300Hz_2b}\includegraphics[width=1.8in,height=2in]{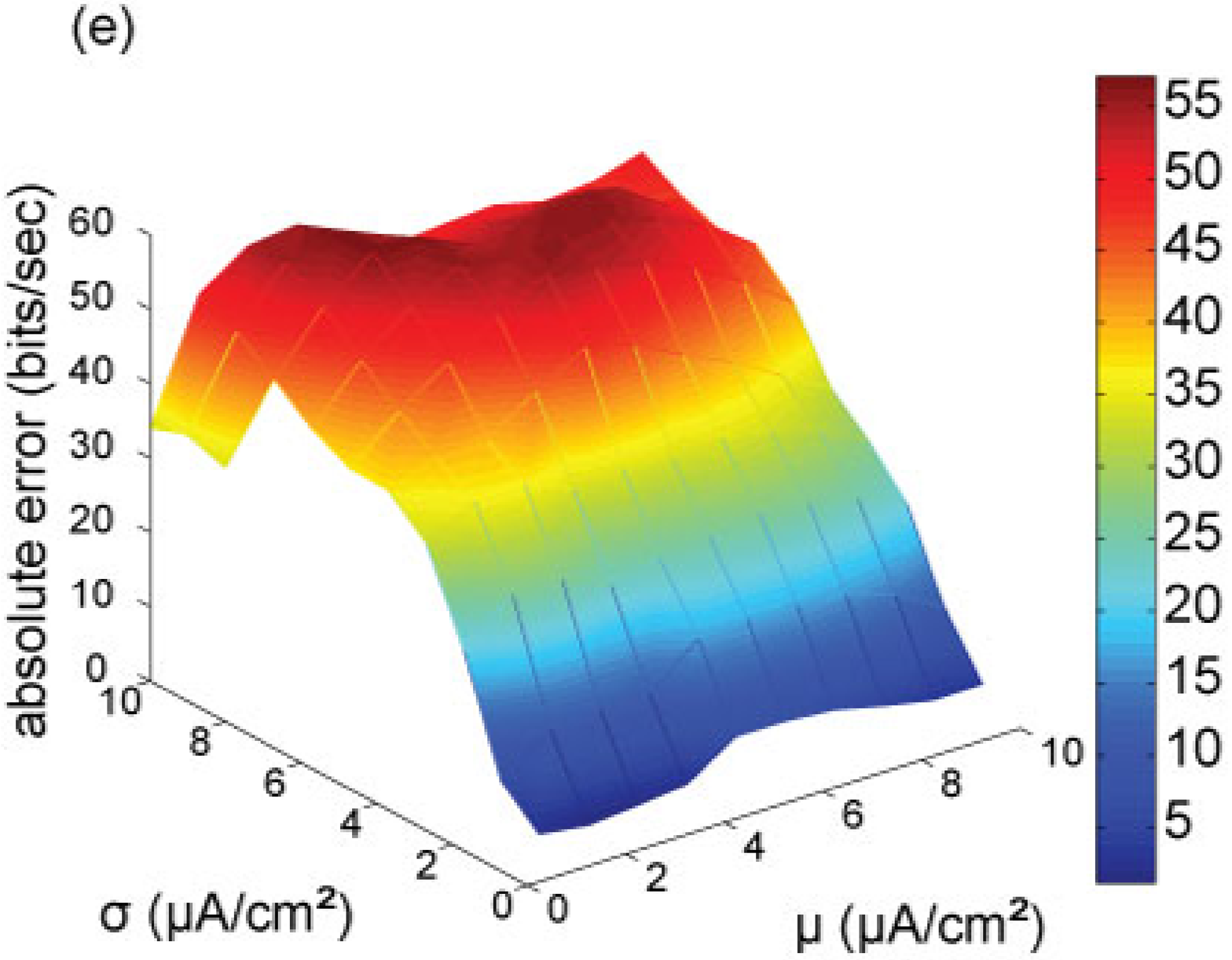}}
\subfigure{\label{fig:digi_300Hz_2c}\includegraphics[width=1.8in,height=2in]{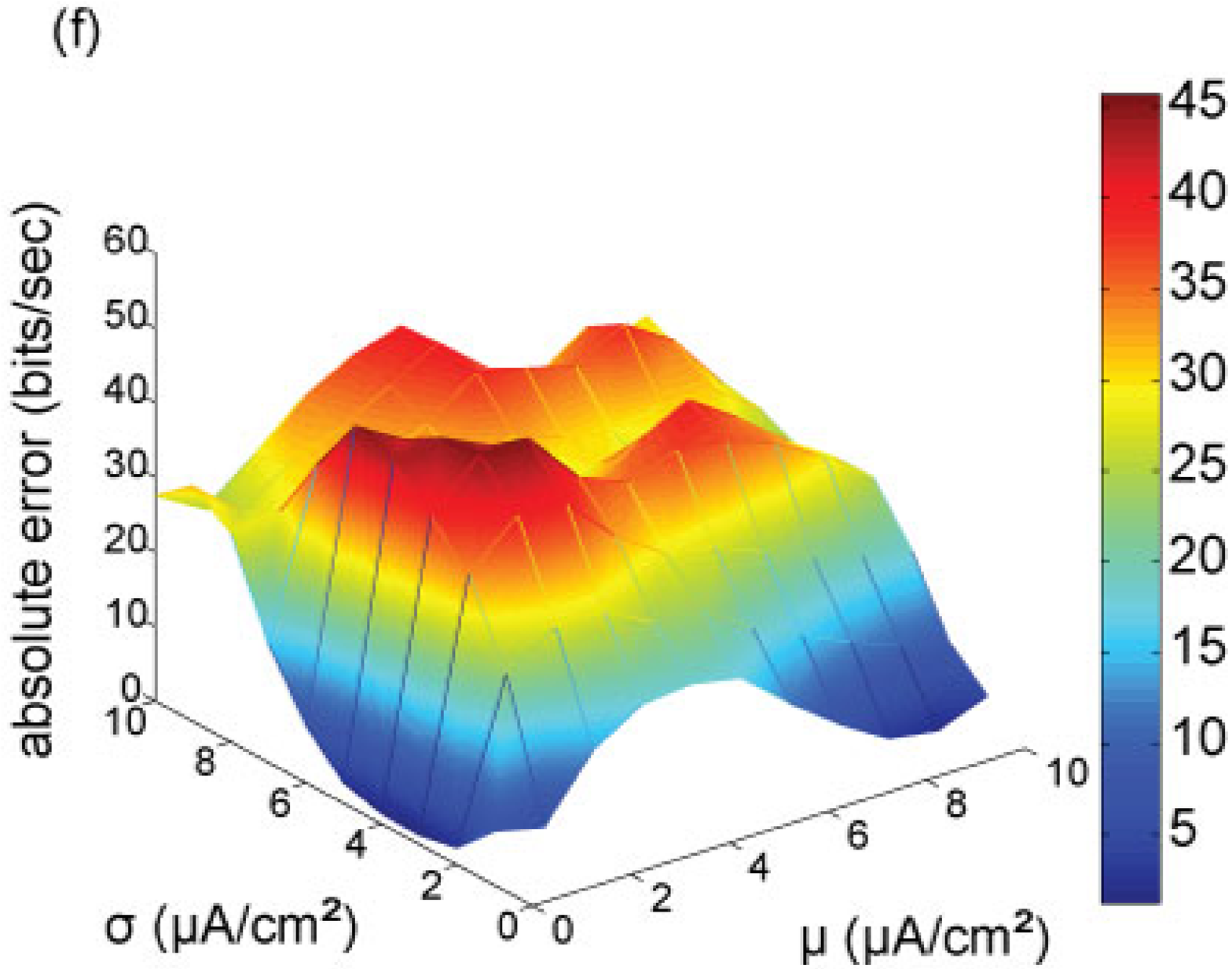}}\\
\caption{(Color online) Langevin model overestimates the information rates in spiking compartments, irrespective of compartment size. The stimulus cut-off frequency was set to 300 Hz. $\mu$ is the mean and $\sigma$ is the  standard deviation of the current injection. Top row: comparison of information rates for models with 3 different areas  (a (1 $ \mu m^{2}$),b ($10 \mu m^{2}$),c (100 $\mu m^{2}$)). Wireframe mesh represents the Langevin implementation of channel noise, while the filled mesh is the Markov implementation. Bottom row:  error surfaces between Langevin and Markov representations  (d (1 $ \mu m^{2}$),e ($10 \mu m^{2}$),f (100 $\mu m^{2}$)).}
\label{fig:spiking_cell}
\end{figure*}

Within a particular size compartment, irrespective of whether the Langevin or Markov model of channel noise was used, the information rate increases with increasing stimulus variance and decreasing mean (Fig. \ref{fig:digi_300Hz_1a}-\ref{fig:digi_300Hz_1c}). In small compartments, information decreases more rapidly than in their larger counterparts as the mean of the stimulus increases. In the largest compartment, increases in stimulus mean had relatively little effect on the information rate. With the highest variance and lowest mean stimuli the Langevin model of channel noise produces information rates of 137, 193 and 262 bits/s in the 1, 10 or 100 $\mu m^{2}$ compartments, respectively. The same stimuli with the Markov model of channel noise produces information rates of 112, 159 and 235 bits/s in the 1, 10 or 100 $\mu m^{2}$ compartments, respectively. Thus, with either model of channel noise, the largest compartment codes approximately $90-110 \ \% $ more information than the smallest compartment when stimulated by low mean and high variance currents.  

In comparison to the Markov model for channel noise, the Langevin model overestimates the information in all compartments irrespective of their size (Fig.  \ref{fig:digi_300Hz_2a}-\ref{fig:digi_300Hz_2c}). For example, when stimulated by low mean and high variance currents the Langevin model overestimates the information by $18 \ \%$, $17 \ \%$ and $10 \ \%$ in the 1, 10 or 100 $\mu m^{2}$ compartments, respectively. Estimates of information rates from compartments with either Langevin or Markov models of channel noise do not tend to converge as the area and, hence, the number of voltage-gated ion channels increases up to 6000 voltage-gated $Na^{+}$ channels and 1800 voltage-gated $K^{+}$ channels in the largest compartment. Subtracting information surfaces obtained using Markov models of channel noise from those obtained using Langevin models showed that the difference in information rates (error surface) is non-linear and strongly dependent upon both the stimulus and the area of the compartment (Fig.  \ref{fig:digi_300Hz_2a}-\ref{fig:digi_300Hz_2c}). The median level of error (ratio of the error surface and the information rate obtained from the Langevin model) between the information rates obtained from the Langevin and Markov models is $ 20 \ \% $, $ 24 \ \% $ and $40 \ \% $ in the 1, 10 or 100 $\mu m^{2}$ compartments, respectively.

\begin{figure*}[htbp]
\centering
\subfigure{\label{fig:frate1}\includegraphics[width=1.8in,height=2in]{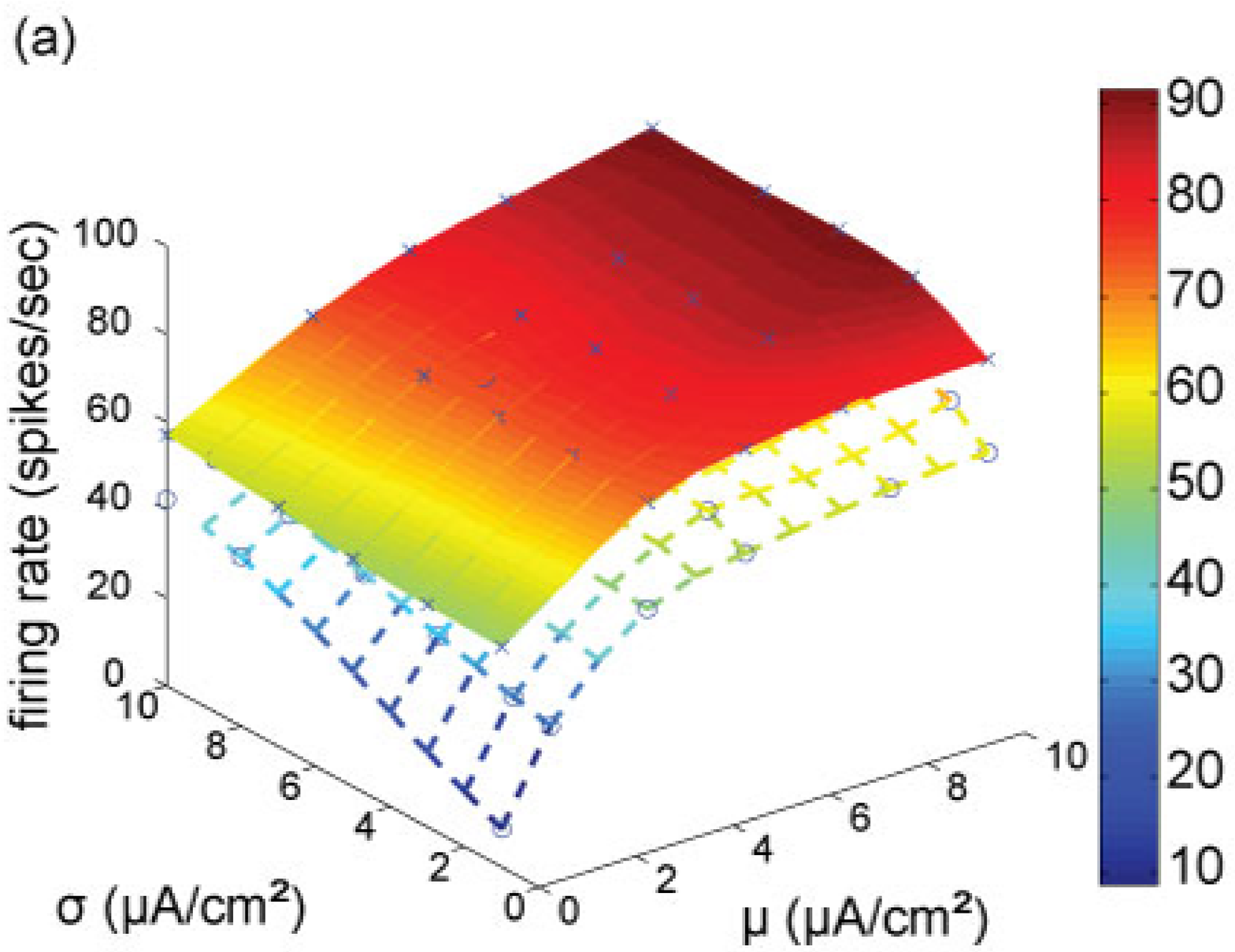}}
\subfigure{\label{fig:frate2}\includegraphics[width=1.8in,height=2in]{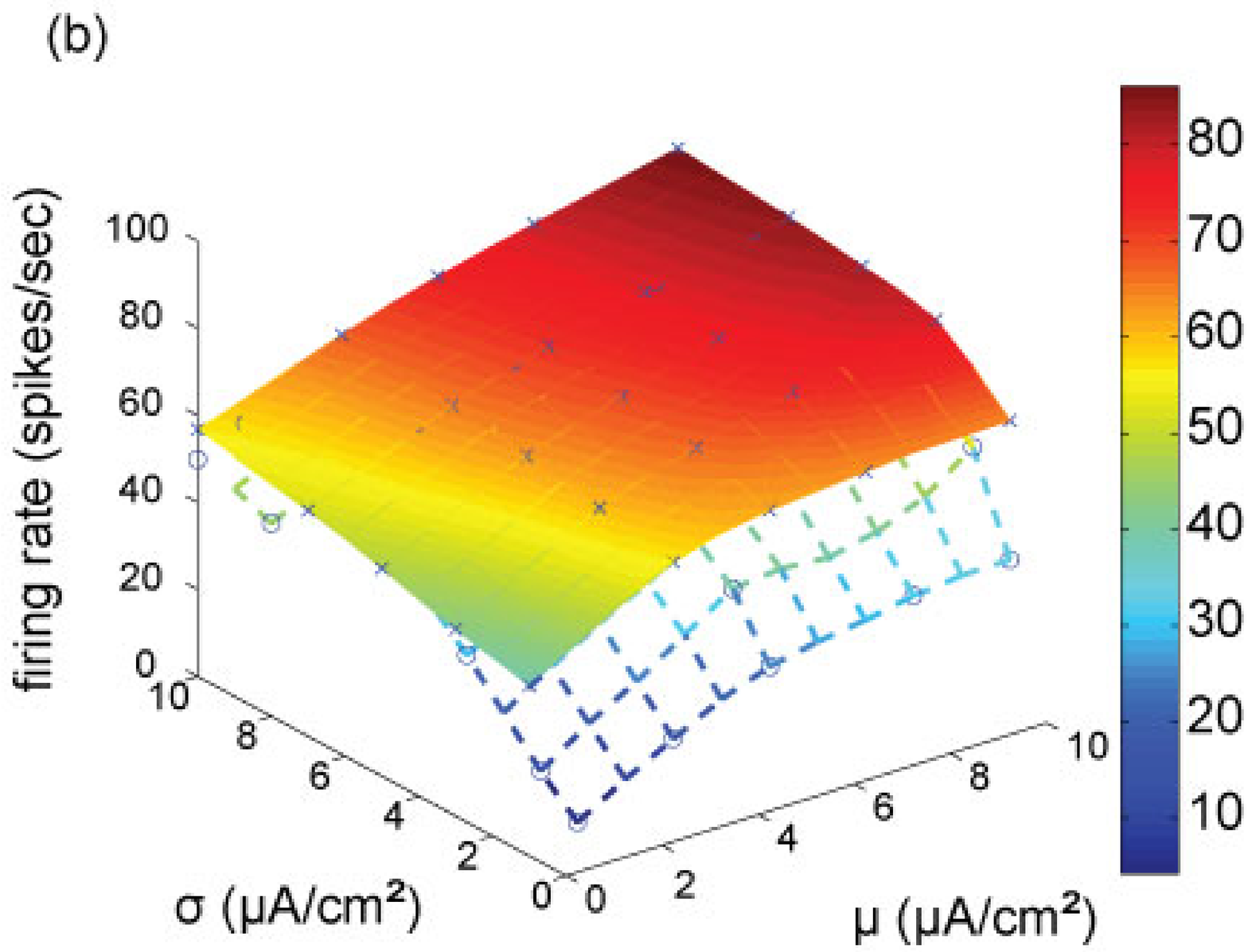}}
\subfigure{\label{fig:frate3}\includegraphics[width=1.8in,height=2in]{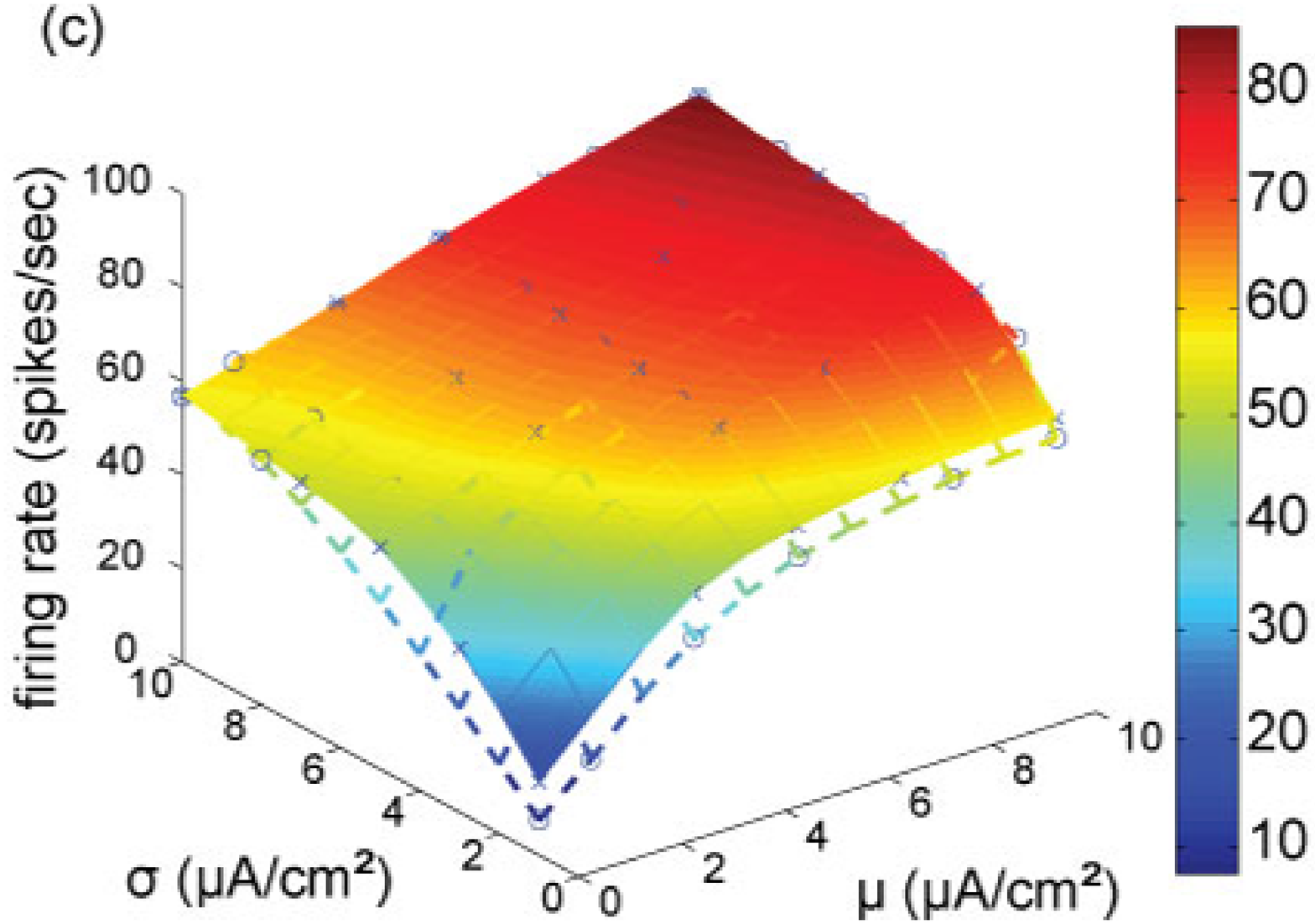}}\\
\subfigure{\label{fig:bps1}\includegraphics[width=1.8in,height=2in]{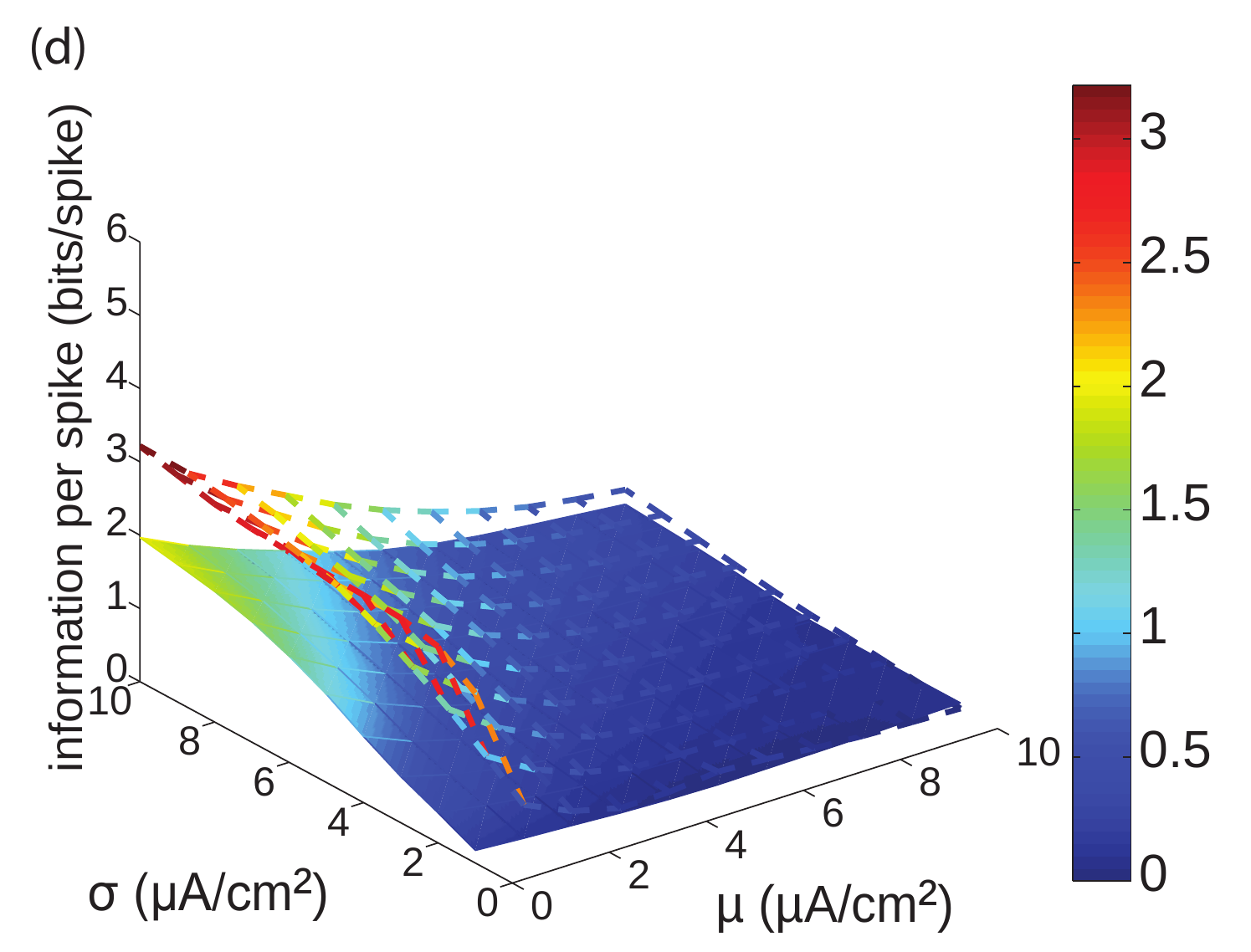}}
\subfigure{\label{fig:bps2}\includegraphics[width=1.8in,height=2in]{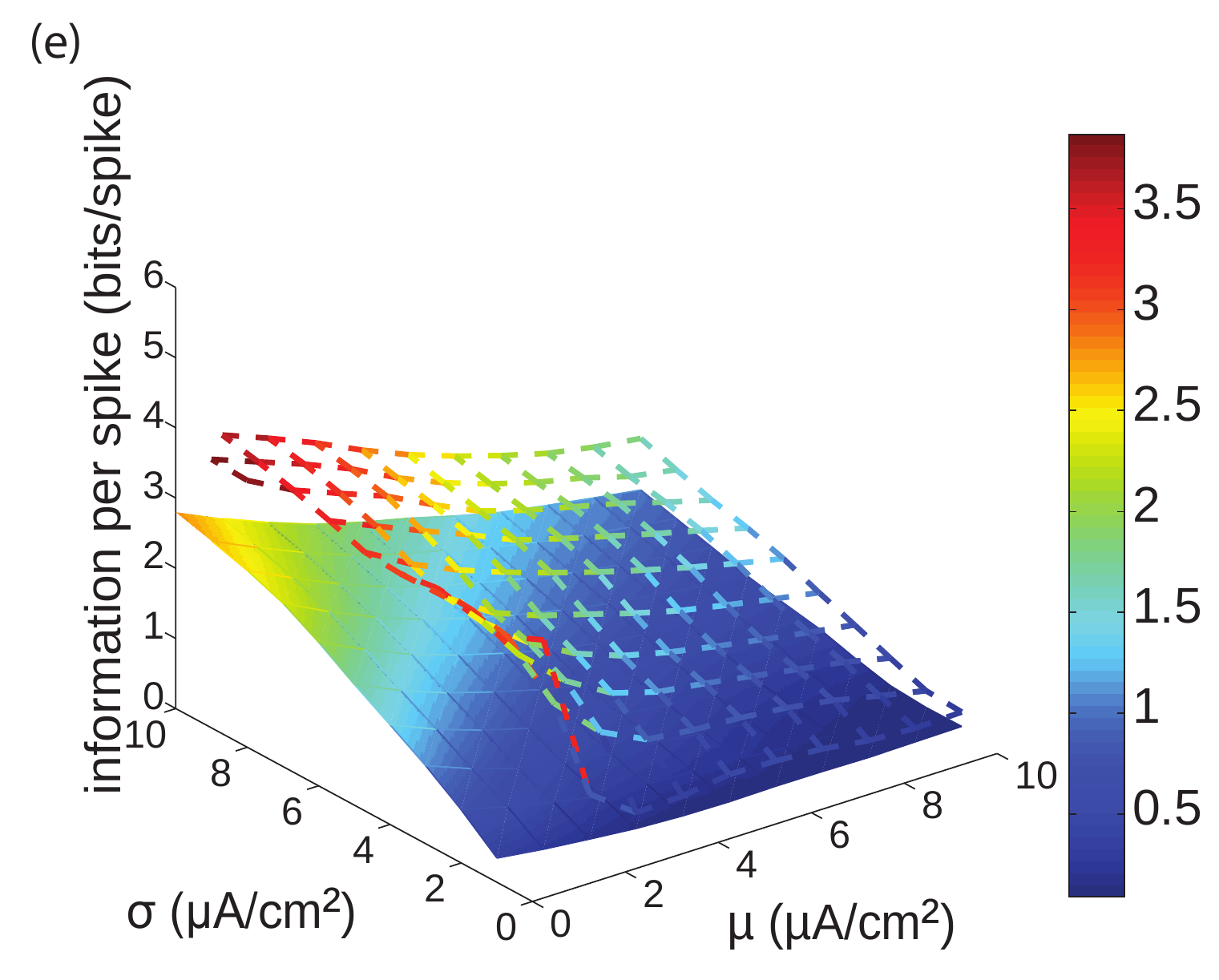}}
\subfigure{\label{fig:bps3}\includegraphics[width=1.8in,height=2in]{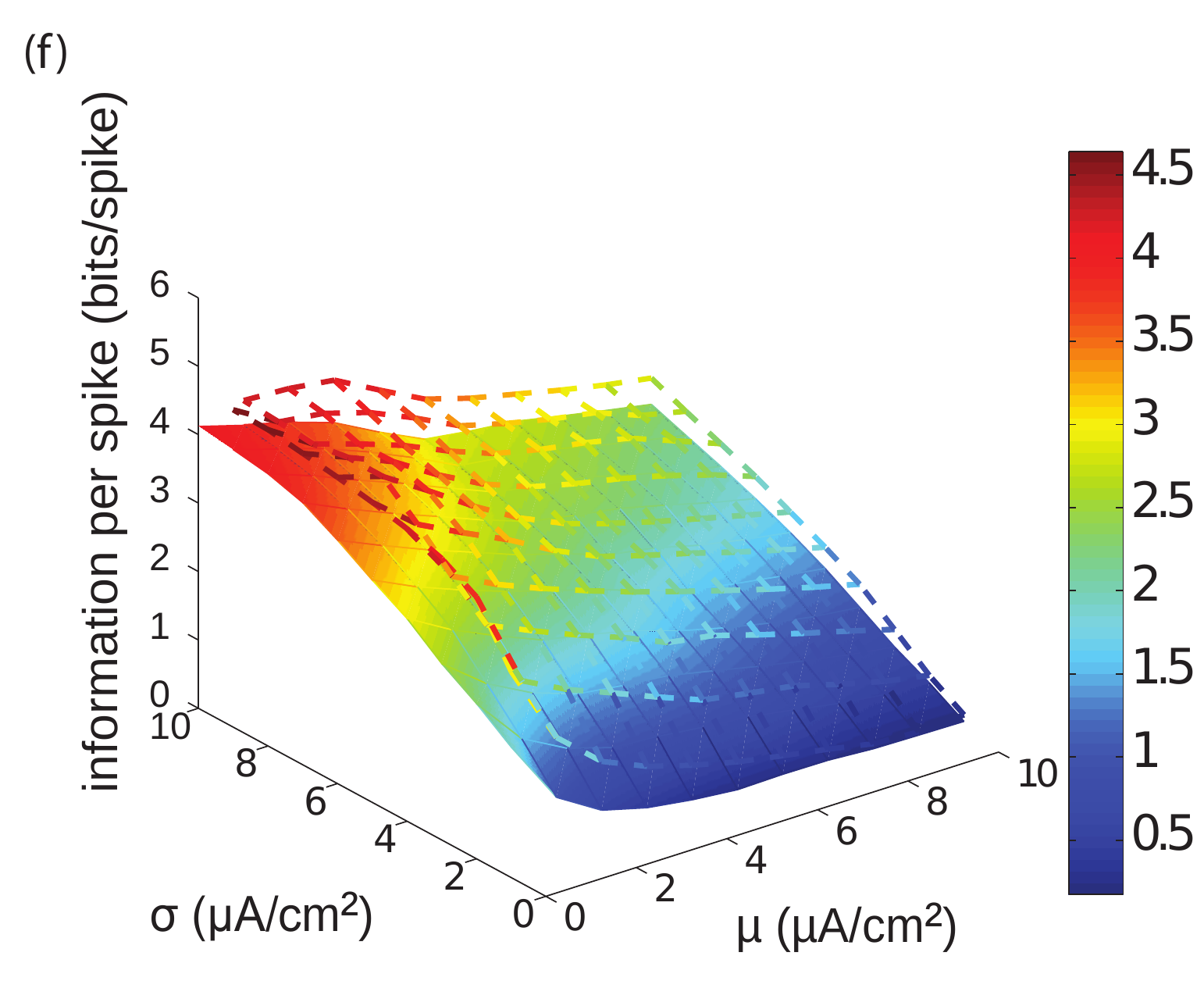}}
\caption{(Color online) Langevin model underestimates the firing rate and overestimates the information obtained per spike of the spiking compartments, regardless of the size of the compartment. Wireframe mesh represents the Langevin implementation of channel noise, while the filled mesh is the Markov implementation. The stimulus cut-off frequency was set to 300 Hz. $\mu$ is the mean and $\sigma$ is the  standard deviation of the current injection. Top row: comparison of firing rate for models with 3 different areas (a (1 $ \mu m^{2}$),b ($10 \mu m^{2}$),c (100 $\mu m^{2}$)). Firing rates are obtained from trial averaged responses to frozen current injections. Bottom row:  comparison of information per spike for models with 3 different areas (d (1 $ \mu m^{2}$),e ($10 \mu m^{2}$),f (100 $\mu m^{2}$)). }
\label{fig:frate}
\end{figure*}

The information rates of compartments that support spikes are dependent on the firing rate as well as the intrinisic and extrinsic noise sources. We calculated the average firing rates of spike trains generated using the frozen white noise current stimuli in the different sized compartments. The Langevin model of channel noise consistently underestimates firing rates in comparison to the Markov model, irrespective of the area of the compartment (Fig. \ref{fig:frate1}-\ref{fig:frate3}). The median underestimation of firing rate by Langevin model is $2 \ \% $ in the 1 $\mu m^{2}$ compartment, $21 \ \% $ in the 10 $\mu m^{2}$ compartment and $36 \ \% $ in the 100 $\mu m^{2}$ compartment. Due to the overestimation of information rates and underestimation of firing rates by the Langevin model, it overestimates the median information coded in each spike relative to the Markov model by $20 \ \% $ in the 100 $\mu m^{2}$ compartment, $42 \ \% $ in the 1 $\mu m^{2}$ compartment and $56 \ \% $ in the 10 $\mu m^{2}$ compartment (Fig. \ref{fig:bps1}-\ref{fig:bps3}). 

\begin{figure*}[htbp]
\begin{center}
\includegraphics[scale=0.4]{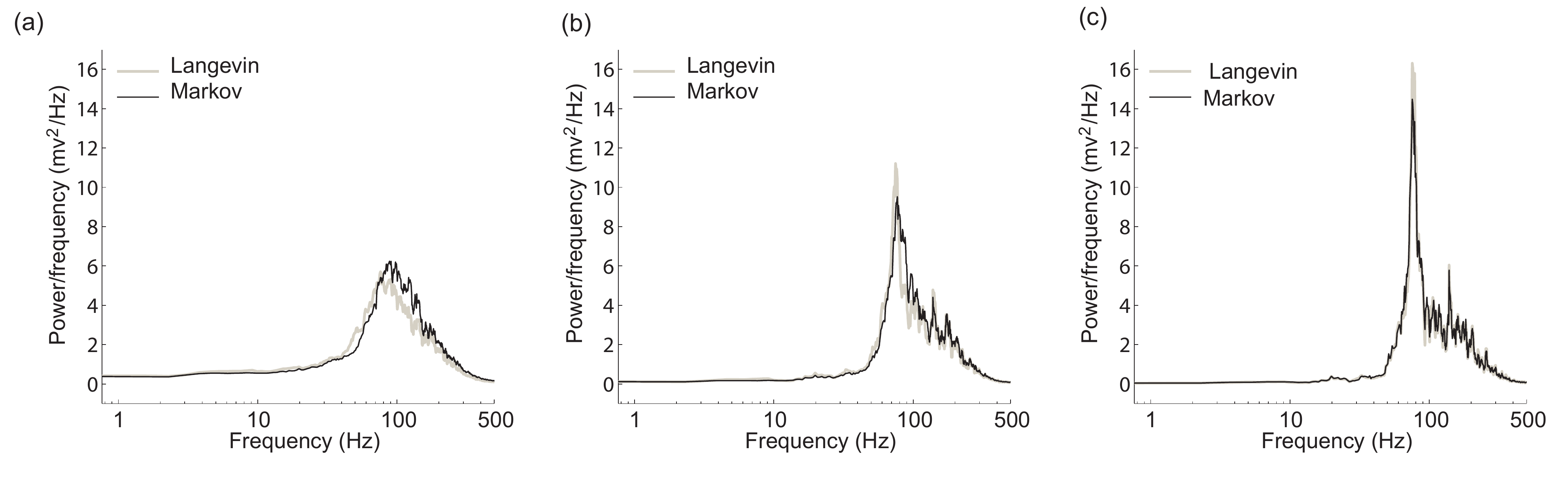}
\caption{(Color online) Comparison of trial averaged, response power spectral density of the voltage traces (with spikes intact) with Langevin and Markov models of channel noise for three different membrane areas (a (1 $ \mu m^{2}$),b ($10 \mu m^{2}$),c (100 $\mu m^{2}$)). Input statistics were sampled from a Gaussian distribution with $\mu = 5 \ \mu A/c{m^2},\sigma=10 \ \mu A/c{m^2} \ and \ \tau_{correlation}=3.3 \ ms$.}
\label{fig:psd_digital}
\end{center}
\end{figure*}

\begin{figure*}[htbp]
\begin{center}
\includegraphics[scale=0.4]{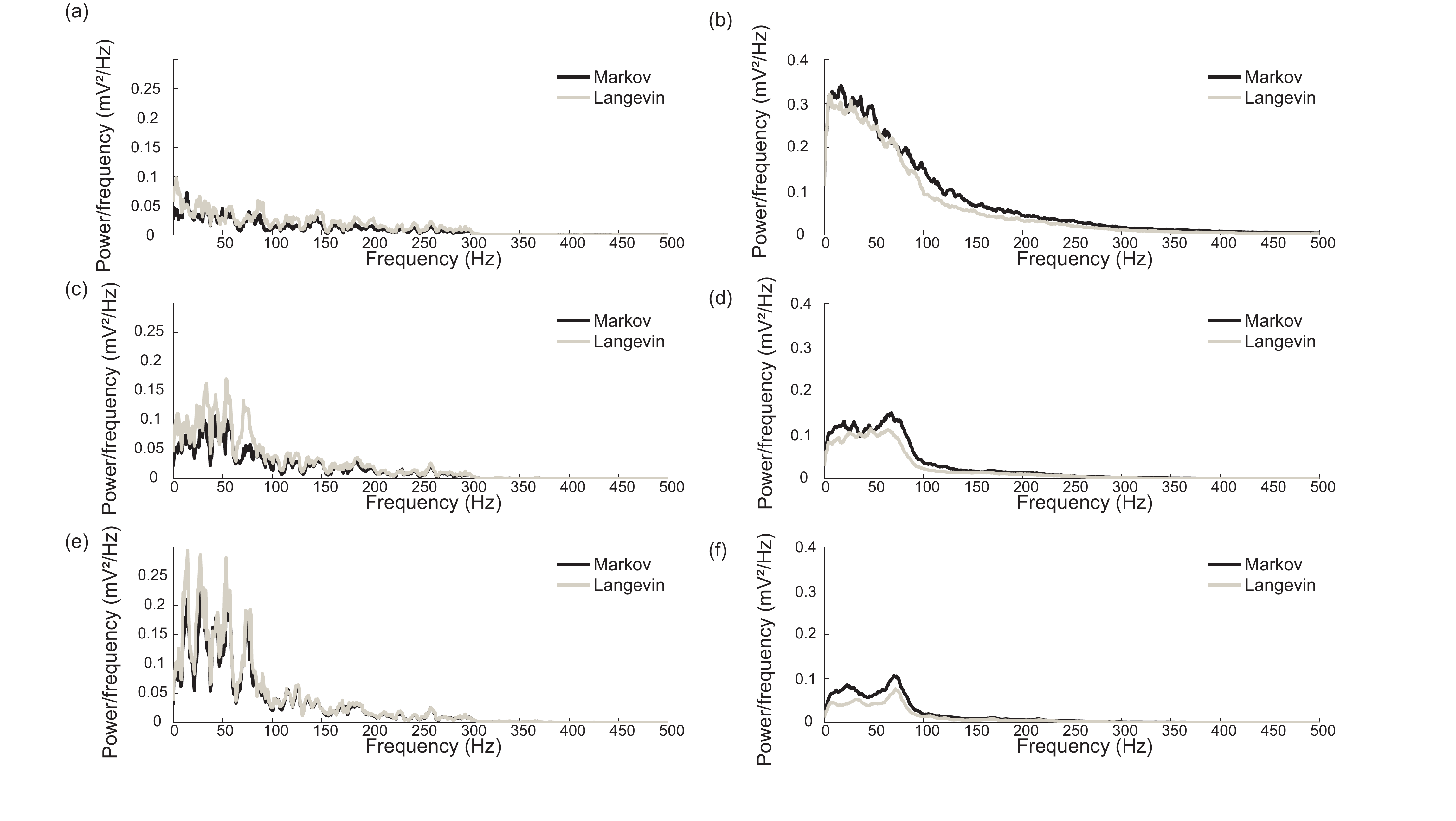}
\caption{(Color online) Comparison of trial averaged signal (a (1 $ \mu m^{2}$),c ($10 \mu m^{2}$),e (100 $\mu m^{2}$)) and noise (b (1 $ \mu m^{2}$),d ($10 \mu m^{2}$),f (100 $\mu m^{2}$)) power spectral density of the voltage traces (with spikes removed) with Langevin and Markov models of channel noise, for three different membrane areas. Input statistics were sampled from a Gaussian distribution with $\mu = 5 \ \mu A/c{m^2},\sigma=10 \ \mu A/c{m^2} \ and \ \tau_{correlation}=3.3 \ ms$.}
\label{fig:psd_digital_SN}
\end{center}
\end{figure*}

Differences in the information rates and firing rates between the Langevin and the Markov models could be due to differences in the distribution and/or frequency content of the voltage responses. We compared the frequency content and distributions of the voltage responses to determine the source of the differences between the models. The frequency content of the voltage signal (including spikes) from these models is similar for a particular sized compartment (Fig. \ref{fig:psd_digital}). We calculated the signal and noise components of the  voltage power spectra after removing the spikes. This was done by removing the voltage waveform between the beginning and end ( $\pm$ 3 ms) of a spike and replacing it with a linear interpolant. After the spikes were removed, the signal was calculated by averaging the voltage responses to a frozen noise stimulus. The noise component was calculated by subtracting the signal from individual responses (see Methods). The signal power in the frequency domain increases as the size of the compartment increases for both the Langevin and Markov models of channel activation. However, as the size of the compartments and, hence the number of voltage-gated ion channels increases, the difference between power spectra from the Langevin and Markov models becomes smaller (Fig. \ref{fig:psd_digital_SN}). In constrast, the average noise power decreases with increasing compartment size and the difference between power spectra from the Langevin and Markov models persists (Fig. \ref{fig:psd_digital_SN}). Thus, the Langevin (approximate) model underestimates the noise, thereby producing an overestimation of information rate in compartments of all sizes. 

\begin{figure*}[htbp]
\begin{center}
\includegraphics[scale=0.4]{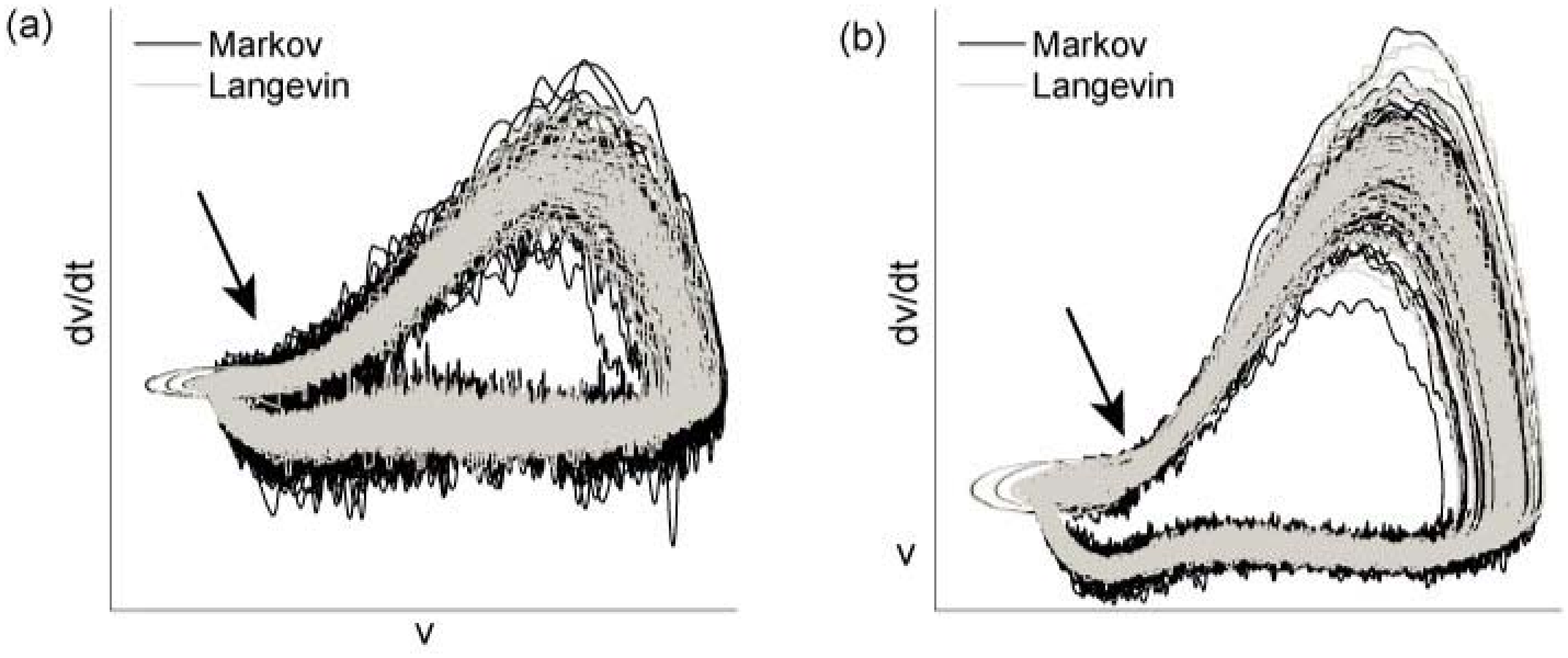}
\caption{(Color online) Phase plots comparing action potential initiation with Langevin and Markov models of channel noise for compartments with 2 different areas (a (10 $ \mu m^{2}$),b ($100 \mu m^{2}$)). Arrow indicates the action potential initiation zone. Input statistics were sampled from a Gaussian distribution with $\mu = 5 \ \mu A/c{m^2},\sigma=10 \ \mu A/c{m^2} \ and \ \tau_{correlation}=3.3 \ ms$.}
\label{fig:phase}
\end{center}
\end{figure*}

The underestimation of noise by the Langevin model may influence spike initiation and thereby the information rate. We constructed phase plots from compartments with either the Langevin or Markov models of channel noise to determine the effect of these models on spike initiation \cite{Izhikevich2006}. These phase plots show that the Langevin based model overestimates the precision of spike initiation in comparison to the Markov model, which introduces a large variance in the timing of spike initiation (Fig. \ref{fig:phase}). Comparison between the phase plots from the 10 $\mu m^{2}$ and 100 $\mu m^{2}$ compartments shows that the differences in spike precision between Langevin and Markov models decreased with increasing compartment size (Fig. \ref{fig:phase}).

\begin{figure*}[htbp]
\begin{center}
\includegraphics[scale=0.4]{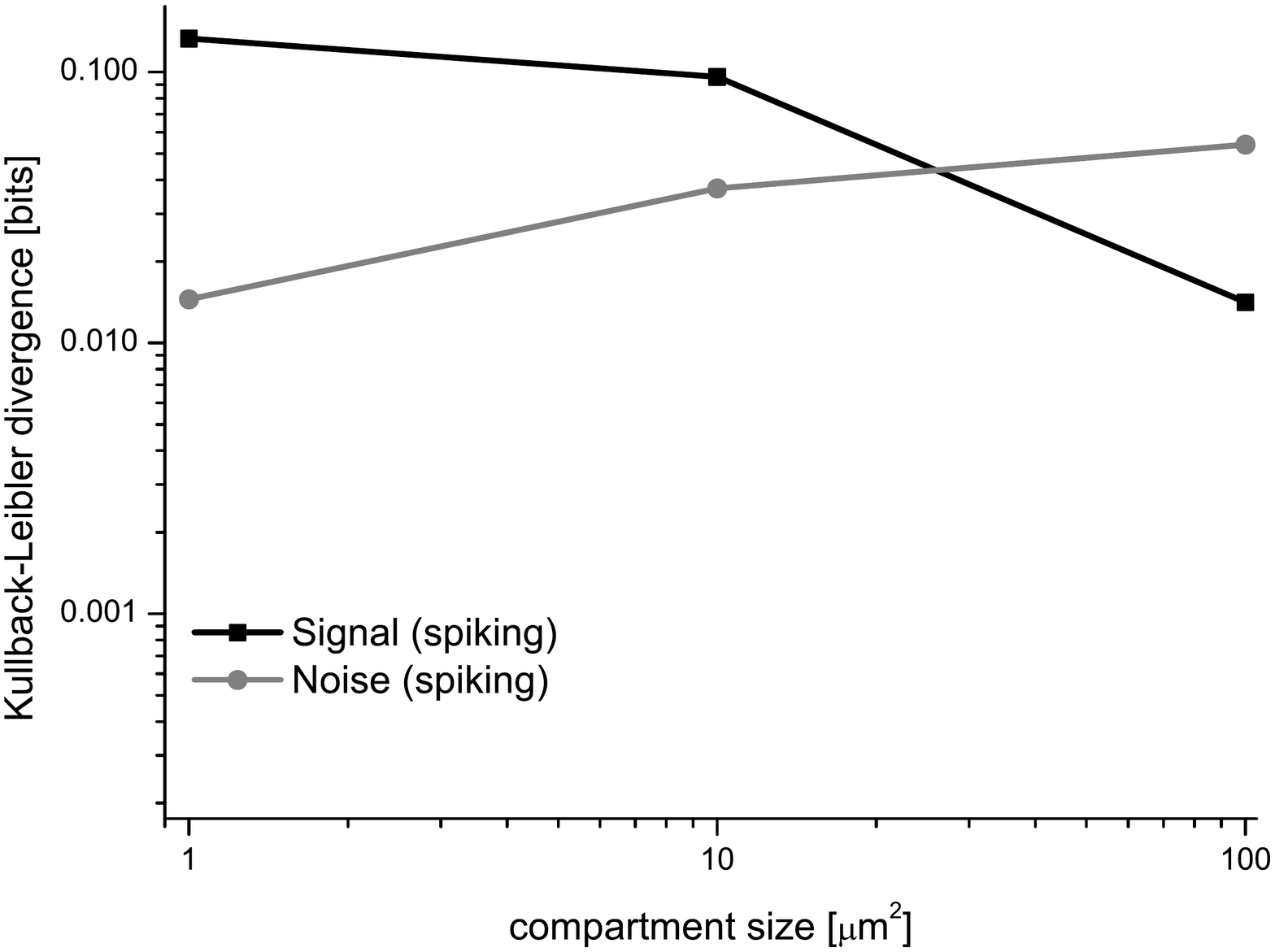}
\caption{(Color online) Comparison of probability density function of Langevin and Markov models of channel noise for spiking compartments. The distance between the distributions is quantified by the Kullback-Leibler (relative entropy) divergence. Input statistics were sampled from a Gaussian distribution with $\mu = 5 \ \mu A/c{m^2},\sigma=10 \ \mu A/c{m^2} \ and \ \tau_{correlation}=3.3 \ ms$.}
\label{fig:pmf_spk}
\end{center}
\end{figure*}

We calculated the differences in the probability density function of the voltage responses produced by the Langevin and Markov models of channel noise to quantify the effect on the voltage responses of the spiking compartments (Fig. \ref{fig:pmf_spk}). The signal and noise probability density functions were calculated after removing the spikes and using a linear interpolant of the voltage in their place. We used the Kullback-Leibler divergence (relative entropy) to quantify the differences between the voltage distributions produced by the Langevin and Markov models \cite{Cover2006}. The relative entropy between two distributions equals zero when they are the same and increases as they diverge. The relative entropy between the signal components of the voltage distributions produced by the Langevin and Markov models decreases with increasing compartment size (Fig. \ref{fig:pmf_spk}). Thus, the signal probability density functions produced by the Langevin and Markov models become more similar in larger compartments. In constrast, the relative entropy between the noise components of the voltage distributions produced by the Langevin and Markov models increases with increasing compartment size (Fig. \ref{fig:pmf_spk}). Thus, the difference between the noise probability density functions produced by the Langevin and Markov models increases in larger compartments. Therefore, although the difference in the signal distribution and frequency content between the Langevin and Markov models drops in larger compartments, differences in the noise distribution and frequency content persist and, in the case of the distribution, even get worse in larger compartments.

\begin{figure*}[htbp]
\begin{center}
\subfigure{\label{fig:SNR2_digi_300Hz_5b}\includegraphics[width=2.8in,height=2in]{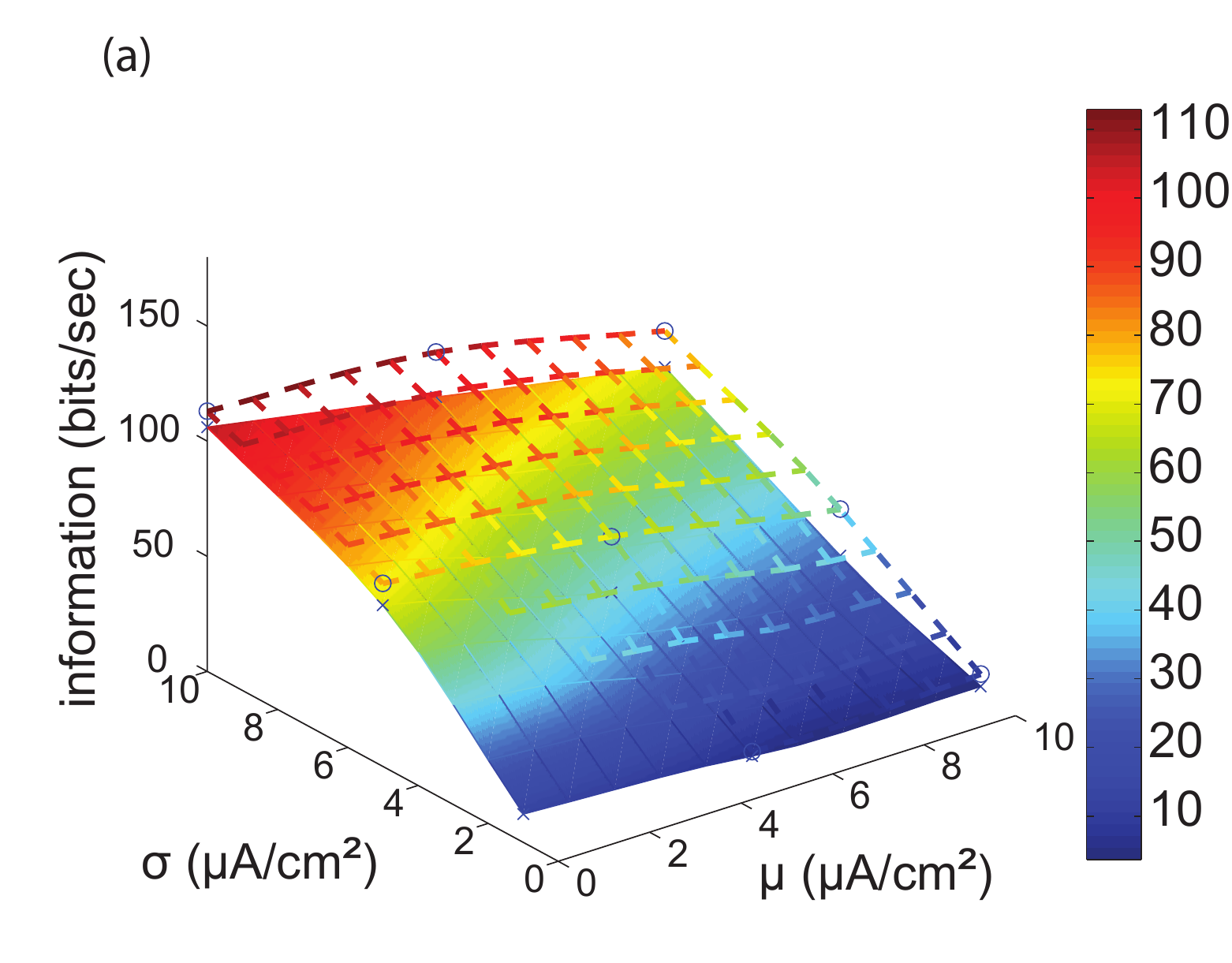}}
\subfigure{\label{fig:SNR20_digi_300Hz_5b}\includegraphics[width=2.8in,height=2in]{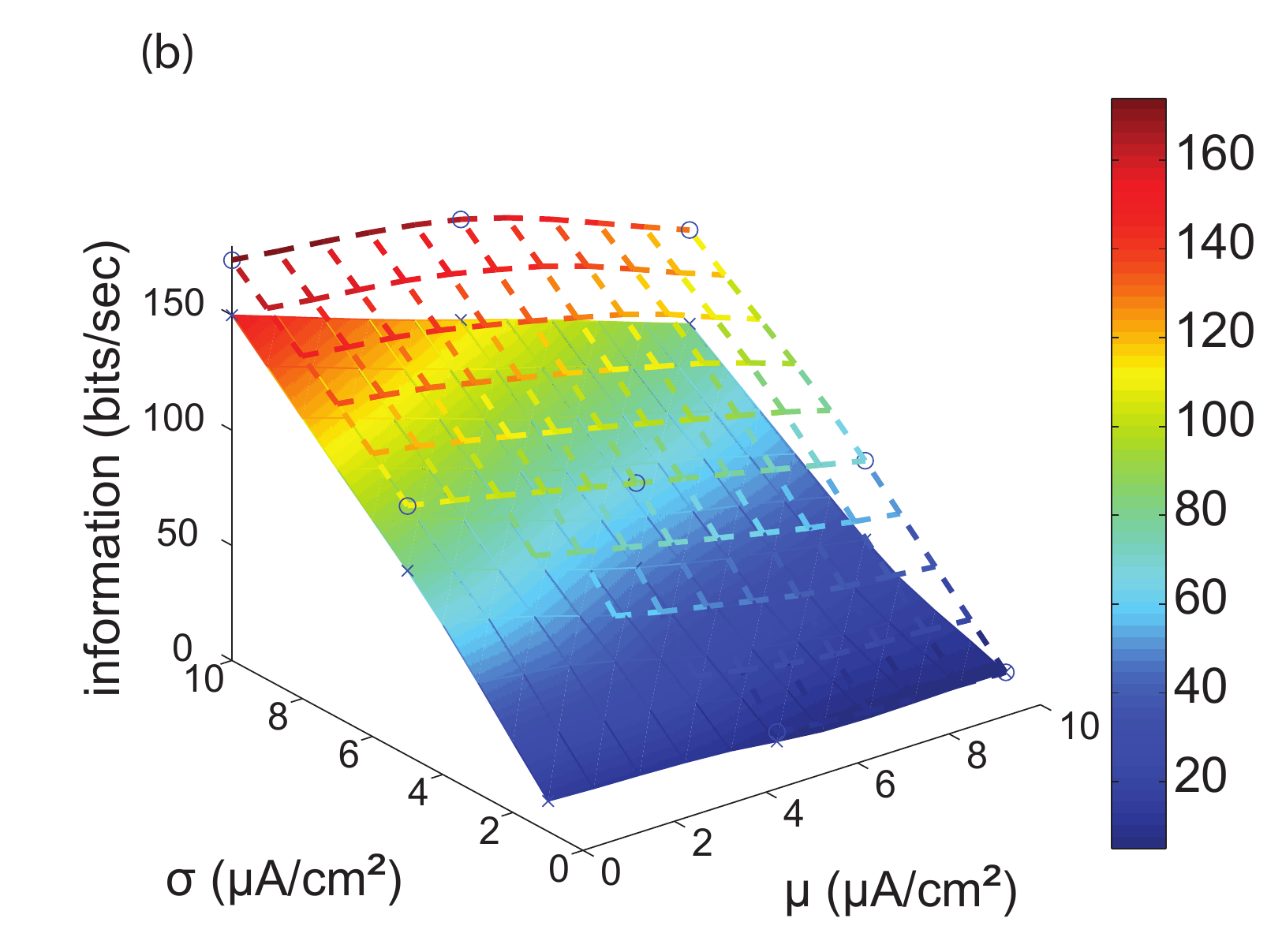}}\\
\caption{(Color online)  Difference between Langevin and Markov models of channel noise with 2 levels of input noise for a $10 \ \mu m^2$ compartment $(f_{c}=300 Hz)$. Filled mesh represents the Markov model, wireframe mesh displays the Langevin model. $\mu$ is the mean and $\sigma$ is the  standard deviation of the current injection. (a): Signal-to-noise ratio equals 2, (b): Signal-to-noise ratio equals 20. }
\label{fig:digiSNR}
\end{center}
\end{figure*}

Signal processing in neurons is constrained not only by intrinsic noise, including channel noise, but also by extrinsic noise in the input stimuli. Extrinsic noise occurs in sensory stimuli, such as photon shot noise, as well as at synapses where there may be variability in the numbers of vesicles released and the number of neurotransmitter molecules they contain \cite{Gerstner2002}. Noise in the input stimulus (extrinsic noise) may affect the extent to which the information rates produced by the Langevin and Markov models differ. We added broad-band Gaussian noise ($\zeta_{noise}(t)$) to the white noise input stimulus to evaluate the role of extrinsic noise on the Langevin and Markov models. Different amounts of noise were added to produce a high or low signal-to-noise ratio (SNR) input stimuli. The addition of extrinsic noise reduced the information rates of the Langevin and Markov models in compartments of all sizes, low SNR stimuli (high extrinsic noise) produces a greater reduction than high SNR stimuli (low extrinsic noise), which produces information rates that approaches those obtained from the noise-free stimuli (Fig. \ref{fig:digiSNR}). The median overestimation of information rates by the Langevin model with low SNR stimuli is $15 \ \% $ in the 1 $\mu m^{2}$ compartment, $26 \ \% $ in the 10 $\mu m^{2}$ compartment and $4 \ \% $  in the 100 $\mu m^{2}$ compartment. With high SNR stimuli the median overestimation of information rates by the Langevin model is $24 \ \% $ in the 1 $\mu m^{2}$ compartment, $36 \ \% $ in the 10 $\mu m^{2}$ compartment and $11 \ \% $  in the 100 $\mu m^{2}$ compartment. The absolute difference between the information rates of the Langevin and Markov models with low SNR stimuli is smaller than with the high SNR stimuli because extrinsic and intrinsic noise variances add. Thus, with low SNR the extrinsic noise is large and adds to the channel noise reducing the overestimation of information rates by the Langevin model. 

\begin{figure*}[htbp]
\centering
\subfigure{\label{fig:ana_300Hz_3a}\includegraphics[width=1.8in,height=2in]{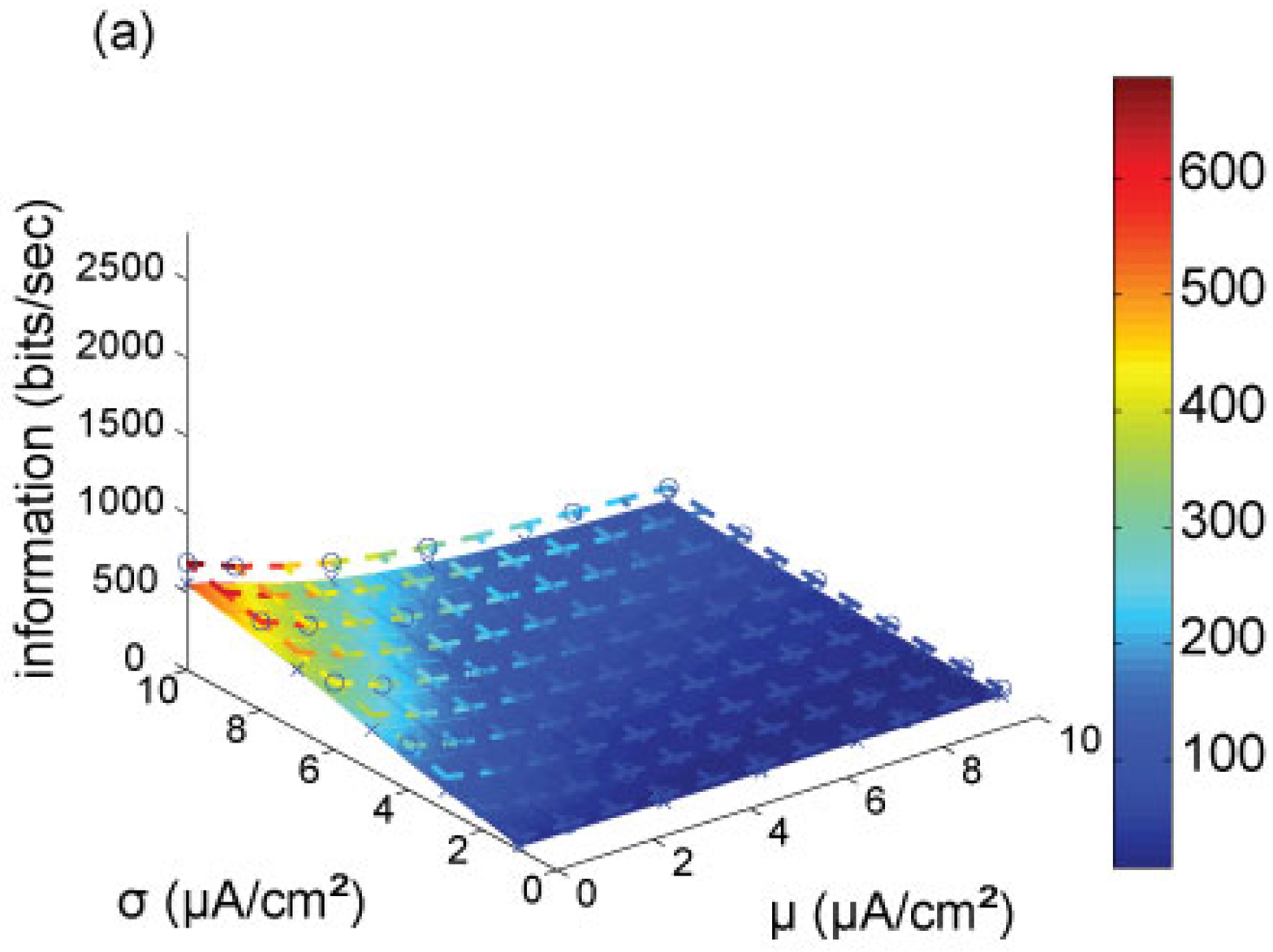}}
\subfigure{\label{fig:ana_300Hz_3b}\includegraphics[width=1.8in,height=2in]{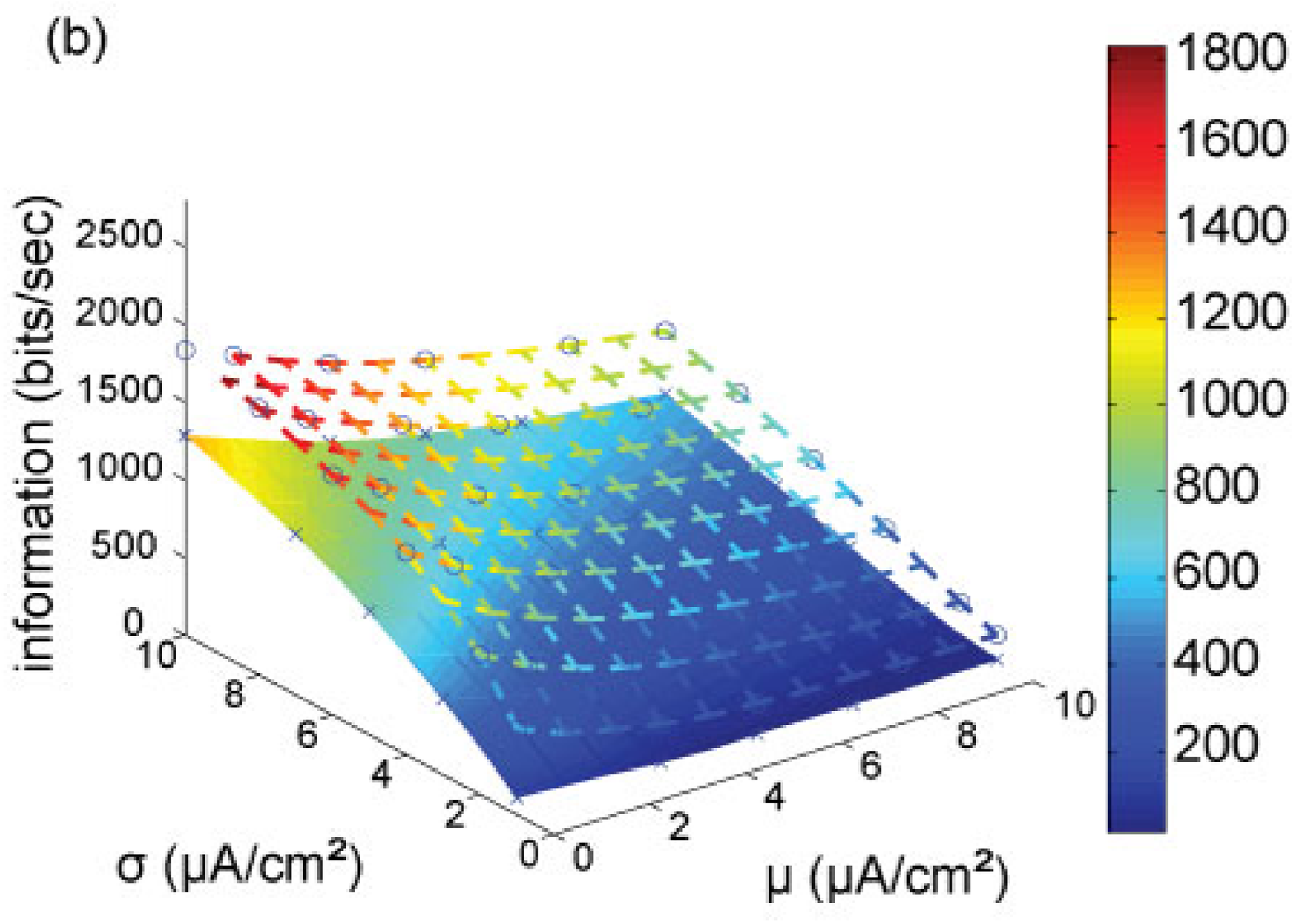}}
\subfigure{\label{fig:ana_300Hz_3c}\includegraphics[width=1.8in,height=2in]{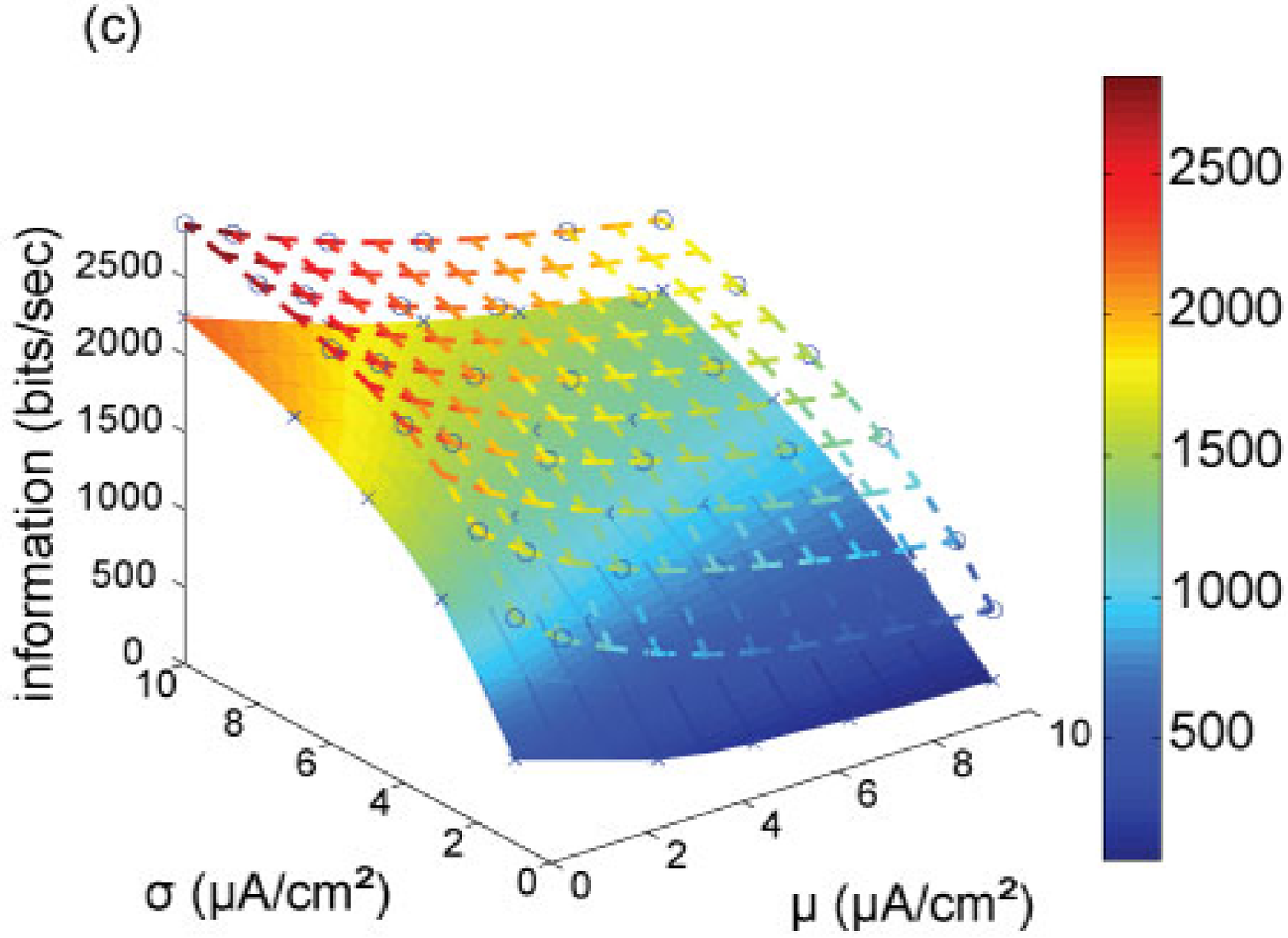}}\\
\subfigure{\label{fig:ana_300Hz_4a}\includegraphics[width=1.8in,height=2in]{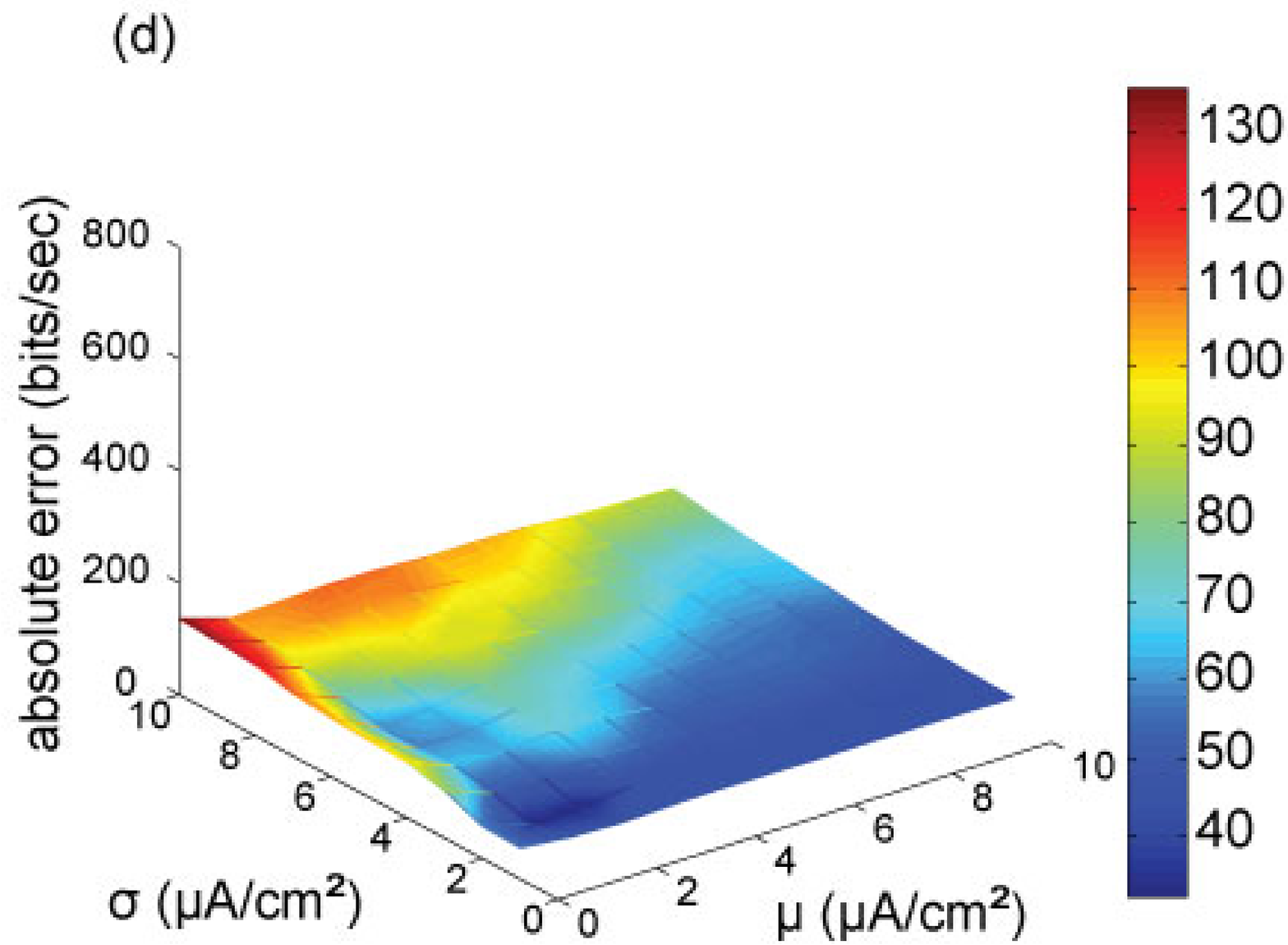}}
\subfigure{\label{fig:ana_300Hz_4b}\includegraphics[width=1.8in,height=2in]{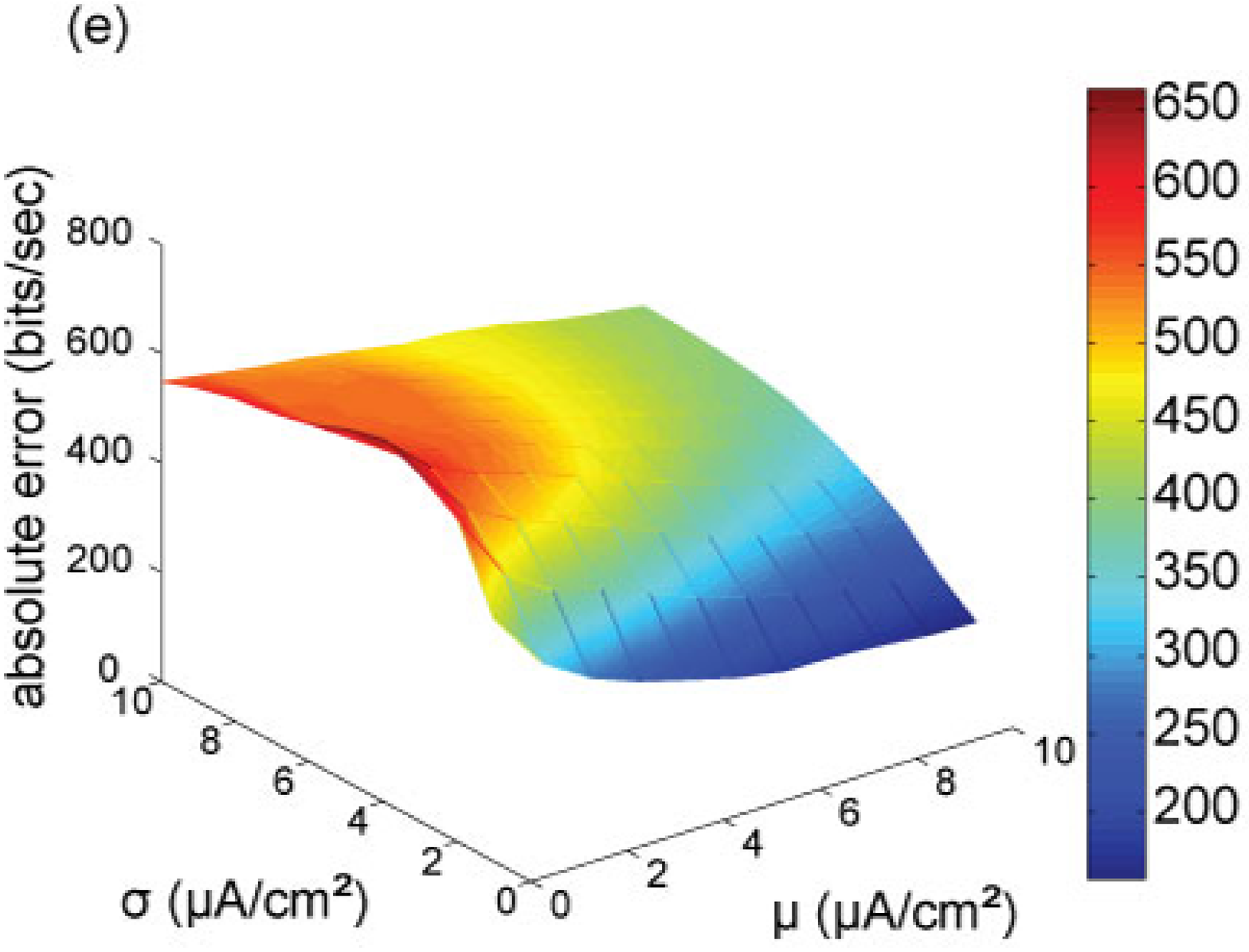}}
\subfigure{\label{fig:ana_300Hz_4c}\includegraphics[width=1.8in,height=2in]{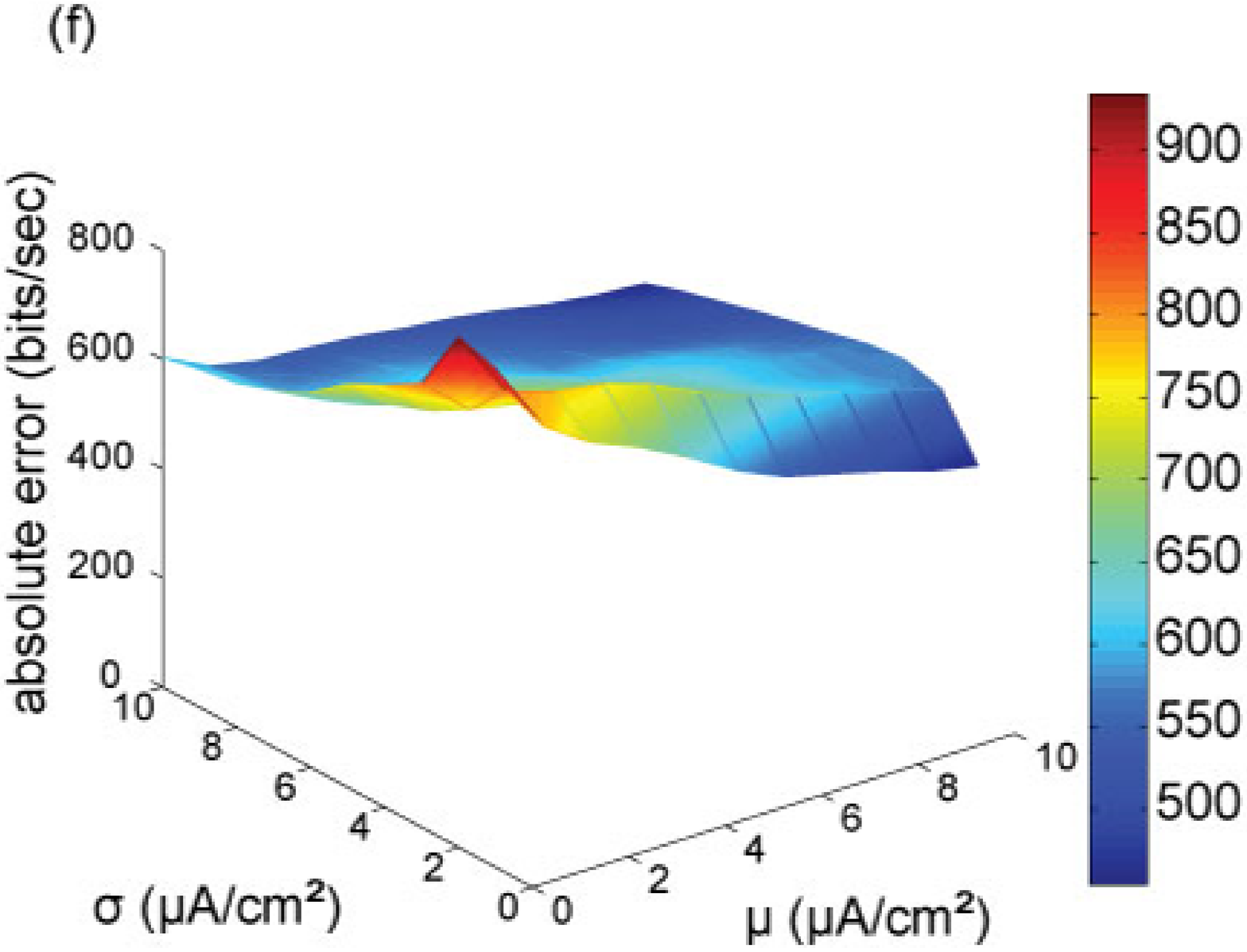}}\\
\caption{(Color online) Langevin model overestimates information rate in  non-spiking compartments. The stimulus cut-off frequency was set to 300 Hz. $\mu$ is the mean and $\sigma$ is the  standard deviation of the current injection. Top row: comparison of information rates for models with 3 different areas  (a (1 $ \mu m^{2}$),b ($10 \mu m^{2}$),c (100 $\mu m^{2}$)).  Wireframe mesh represents the Langevin implementation of channel noise, while the filled mesh is the Markov implementation. Bottom row:  error surfaces between Langevin and Markov representations  (d (1 $ \mu m^{2}$),e ($10 \mu m^{2}$),f (100 $\mu m^{2}$)). }
\label{fig:analog_cell}
\end{figure*}

We also compared Langevin and Markov formulations of channel noise using non-spiking single compartment models, possessing only voltage-gated $K^{+}$ channels and additional leak conductances. As with the spiking compartments above we simulated the responses of these models to low-pass filtered Gaussian signals with different means and variances. Again, the area of each compartment model was either 1, 10 or 100 $\mu m^{2}$, within these compartments the specific density of the voltage-gated $K^{+}$ channels was $18  /  \mu  m^2$. The highest information rates in non-spiking compartments are obtained from input stimuli with low means and high variances (Fig. \ref{fig:ana_300Hz_3a}-\ref{fig:ana_300Hz_3c}). With the highest variance and lowest mean stimuli the Langevin model of channel noise produces information rates of 687, 1833 and 2844 bits/s in the 1, 10 or 100 $\mu m^{2}$ compartments, respectively. The same stimuli with the Markov model of channel noise produce information rates of 551, 1284 and 2240 bits/s in the 1, 10 or 100 $\mu m^{2}$ compartments, respectively. Thus, with either model of channel noise, the largest compartment codes approximately $300 \ \% $ more information than the smallest compartment when stimulated by low mean and high variance currents.   

The Langevin model overestimates the information rates in non-spiking compartments in comparison to the Markov model of channel noise (Fig.\ref{fig:ana_300Hz_3a}-\ref{fig:ana_300Hz_3c}). For example, when stimulated by low mean and high variance currents the Langevin model overestimates the information by $ 20 \ \% $, $ 29 \ \% $ and $ 21 \ \% $ in the 1, 10 or 100 $\mu m^{2}$ non-spiking compartments, respectively. The difference (error surface) between the Langevin and the Markov models for a particular size non-spiking compartment is smoother than in spiking compartments (Fig. \ref{fig:ana_300Hz_4a}-\ref{fig:ana_300Hz_4c}) and increases with increasing compartment size. As in the spiking compartments, these error surfaces are non-linear. The median overestimation of the information rates obtained from the Langevin model in comparisons to the Markov model is $47 \ \% $, $52 \ \% $ and $33 \ \% $ in the 1, 10 or 100 $\mu m^{2}$ non-spiking compartments, respectively. Thus, the overestimation of information rates in non-spiking compartments by the Langevin model relative to the Markov model is greater in non-spiking than in spiking compartments.

\begin{figure*}[htbp]
\begin{center}
\includegraphics[scale=0.4]{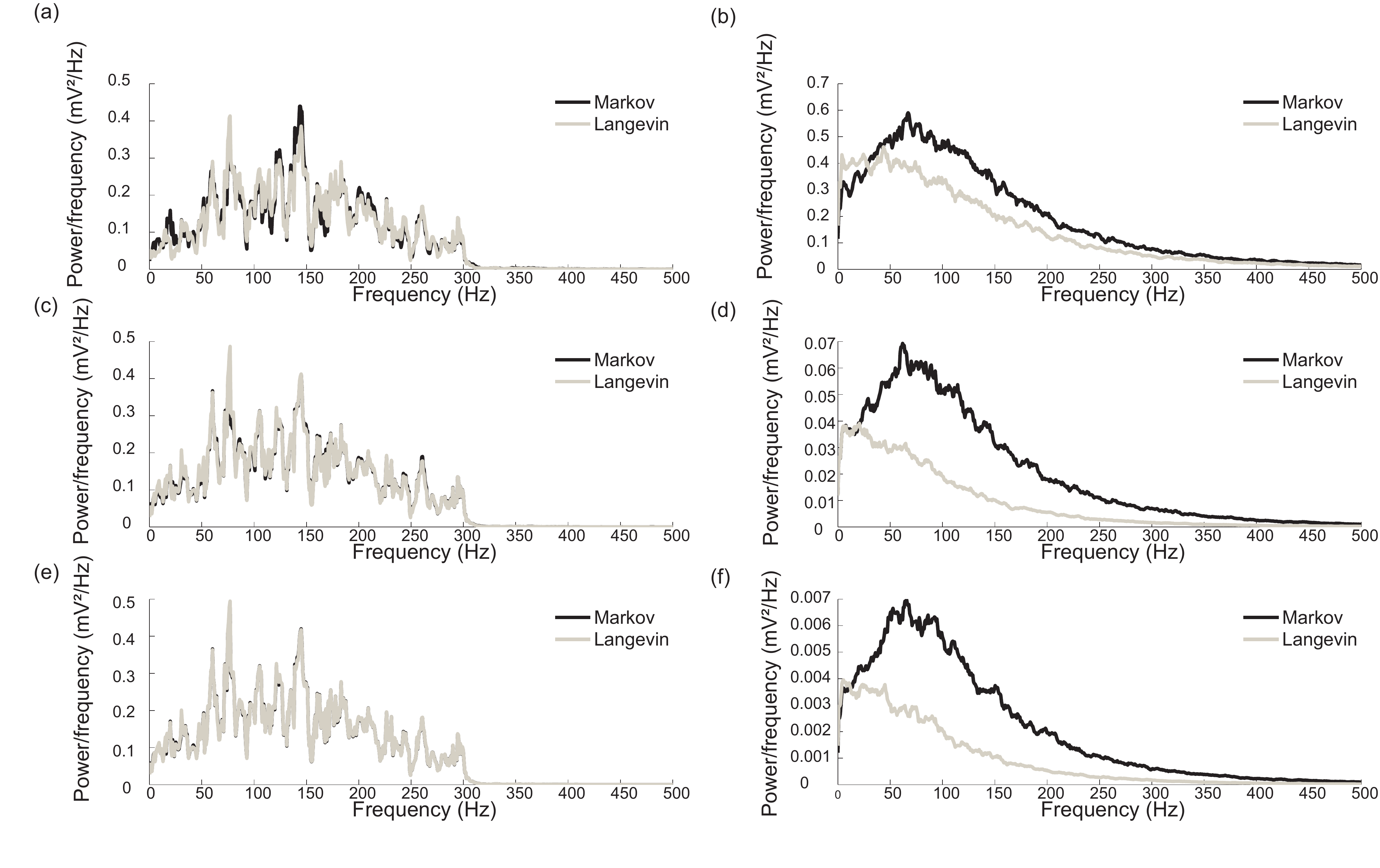}
\caption{(Color online) Comparison of trial averaged signal (a (1 $ \mu m^{2}$),c ($10 \mu m^{2}$),e (100 $\mu m^{2}$)) and noise (b (1 $ \mu m^{2}$),d ($10 \mu m^{2}$),f (100 $\mu m^{2}$))  power spectral density of the voltage traces with Langevin and Markov models of channel noise, for three different membrane areas. Input statistics were sampled from a Gaussian distribution with $\mu = 5 \ \mu A/c{m^2},\sigma=10 \ \mu A/c{m^2} \ and \ \tau_{correlation}=3.3 \ ms$.}
\label{fig:psd_anal_SN}
\end{center}
\end{figure*}

We compared the frequency content and distributions of the voltage responses to determine the source of the differences between the models. The signal and noise components of the voltage responses of the non-spiking compartments were calculated as described previously for the spiking compartments (see Methods). Signal power in non-spiking compartments is independent of compartment size and is similar for both Langevin and Markov models of channel noise (Fig. \ref{fig:psd_anal_SN}). In contrast, the Langevin model underestimates noise power in comparison to the Markov model, the difference between these models increasing as the size of the compartment increases. Thus, as in spiking compartments, the Langevin model underestimates noise producing an overestimation of the information rate in non-spiking compartments of all sizes (Fig. \ref{fig:psd_anal_SN}). 


\begin{figure*}[htbp]
\begin{center}
\includegraphics[scale=0.4]{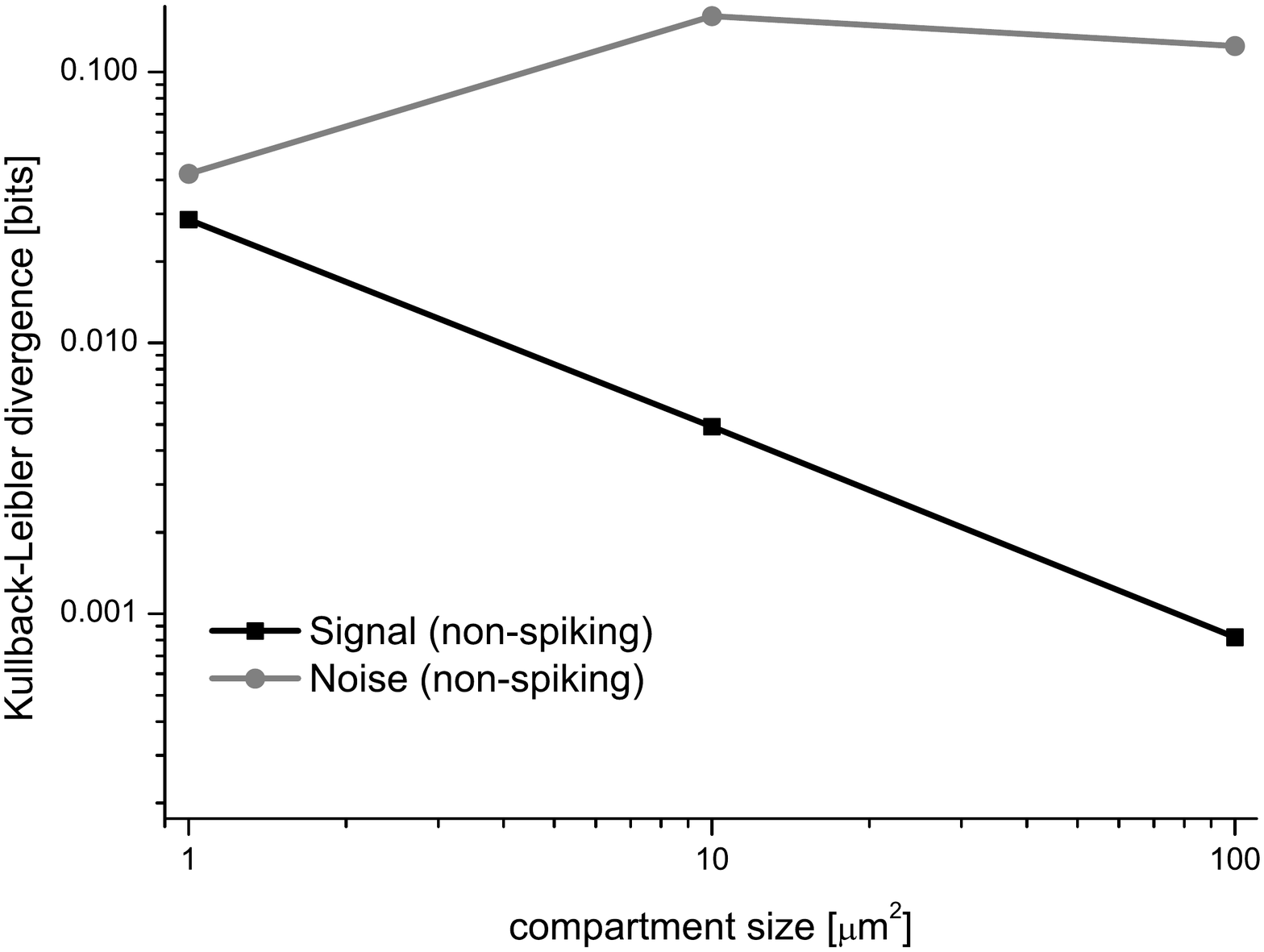}
\caption{(Color online) Comparison of probability density function of Langevin and Markov models of channel noise for non-spiking compartments. The distance between the distributions is quantified by the Kullback-Leibler (relative entropy) divergence. Input statistics were sampled from a Gaussian distribution with $\mu = 5 \ \mu A/c{m^2},\sigma=10 \ \mu A/c{m^2} \ and \ \tau_{correlation}=3.3 \ ms$.}
\label{fig:pmf_ana}
\end{center}
\end{figure*}

We also calculated the differences in the probability density function of the voltage responses produced by the Langevin and Markov models of channel noise to quantify their effect on the voltage responses of the non-spiking compartments (Fig. \ref{fig:pmf_ana}). As in the spiking compartments, we used the Kullback-Leibler divergence (relative entropy) to discriminate the distribution produced by the Langevin model from that of the Markov model \cite{Cover2006}. The relative entropy between the signal components of the voltage distributions produced by the Langevin and Markov models decreases with increasing compartment size (Fig. \ref{fig:pmf_ana}). Thus, the signal probability density functions produced by the Langevin and Markov models become more similar in larger compartments. However, the relative entropy between the noise components of the voltage distributions produced by the Langevin and Markov models increases with increasing compartment size (Fig. \ref{fig:pmf_ana}). Thus, the difference between the noise probability density functions produced by the Langevin and Markov models increases in larger non-spiking compartments. Therefore, as in spiking compartments, although the difference in the distribution of the voltage signal produced by the Langevin and Markov models decreases with increasing compartment size the difference between the distribution of the voltage noise increases with increasing compartment size.

\section{\label{sec:discuss}Discussion}

Our simulations show that Langevin (approximate) models for channel noise produce higher estimates of information rates than Markov (exact) models under all the conditions that were tested. Nevertheless, information rate surfaces estimated by the Langevin models are qualitatively similar to those estimated by the Markov models. Although we expected that differences between exact and approximate methods would be more pronounced in smaller compartments with fewer voltage-gated ion channels \cite{White2000}, our simulations show that Langevin models overestimate information rates irrespective of the size of the compartment. The overestimation of information rates by Langevin models applies to spiking as well as non-spiking compartments. However, with deteriorating input signal quality, the difference in the information rate estimates between the Langevin and Markov models decreases because the variances of the extrinsic and intrinsic noise sources add to reduce the underestimation by the Langevin channel noise model.

The overestimation of information rates by the Langevin model, when compared to those estimated by the Markov model, is due to differences in both the frequency content and voltage distribution of the signal and noise in spiking and non-spiking compartments. The distribution of the signal becomes more similar as compartment size increases. Likewise, signal power becomes more similar at each frequency as compartment size increases. In contrast to the signal, the distribution of the noise becomes less similar as compartment size increases and the power of the noise in the Langevin model is lower at most frequencies in comparison to that of the Markov model. Underestimation of the noise power in the Langevin model causes an overestimation of SNRs and, hence, information rates even in the largest compartments we simulated. 

Using pulse and step inputs, Bruce (2009) reported that the differences between Langevin and Markov models are due to the underestimation of the variance of the number of open channels by the Langevin method \cite{Bruce2009}. For the Langevin model to reproduce exactly the same distribution of spike initiations as the Markov model, the perturbation of the activation and the inactivation variables should be non-Gaussian and correlated \cite{Bruce2009}. Additionally, voltage-gated ion channels possess one or more particles that contribute to the probability of the opening and closing of the channel's gating mechanism \cite{HODGKIN1952}. The combined behavior of multiple activation particles ($m^{3}$ and/or $n^{4}$) is not captured by the Langevin model, although it converges to the exact Markov model with only a single activation particle per channel ($m$ and/or $n$). Thus, the main reason that the Langevin model underestimates noise and overestimates information rates in our simulations is that it underestimates the variability in the numbers of open channels. This is not a problem when using Markov models of channel noise, which infer the exact number of ion channels opened during each time step \cite{Fall2002}. Hence, the Langevin model produces higher information rates than those obtained using the Markov model for channel noise. Irrespective of whether the response is spiking or non-spiking, this underestimation of the noise due to the presence of multiple activation particles ($m^{3}$ and/or $n^{4}$) may be remedied by analytical derivation of a correction factor for the Langevin equation \cite{Bruce2009}.  

Calculations based on electrophysiological recordings from spiking neurons suggest that their information rates may reach approximately 300 bits $s^{-1}$ \cite{Roddey1996}. The highest information rates in our simulations likewise reach approximately 250 bits $s^{-1}$ in the largest spiking compartment though precise comparison is not possible due to differences in the size and channel composition of the experimentally recorded neurons, additional noise sources not incorporated into our simulations, the stimulus used and the temperature at which the experimental responses are recorded. Calculations based on electrophysiological recordings from non-spiking neurons typically produce higher information rates than those from spiking neurons. In non-spiking neurons, such as photoreceptors or large monopolar cells in insect retina, information rates may reach approximately 1500 bits $s^{-1}$ \cite{RuytervanSteveninck1996}. Our simulations of non-spiking compartments also have higher information rates than similarly sized spiking compartments. Recent work in fly retina has also shown experimentally that information rates are dependent upon photoreceptor size, the smaller photoreceptors having information rates of approximately 200 bits $s^{-1}$ and larger photoreceptors information rates of approximately 1200 bits $s^{-1}$ \cite{Niven2007}. Our simulations produce information rates of approximately 2500 bits $s^{-1}$ in the largest non-spiking compartments. Again, direct comparison of information rates from our simulations with those from experiments is difficult because of differences in size and channel composition of the experimentally recorded neurons, additional noise sources not incorporated into our simulations, the stimulus used and the temperature at which the experimental responses are recorded. Nevertheless, comparison with experimental information rates suggests that both our spiking and non-spiking simulations are operating within a biologically plausible range. 

In our simulations, we assume that the probability of switching between states depends only on the present state of a channel and not on the history of previous states a channel has occupied or the duration of the time that a channel has remained in a particular state. Experimental evidence in mouse Leydig cells \cite{Bandeira2008} and locust extensor tibiae muscle \cite{Fulinski1998} shows that channel noise in some ion channels (BK channels) is non-Gaussian and non-Markovian suggesting that both Langevin and Markov models of noise are themselves only approximate representations of channel noise in neural systems. To account for non-Gaussian and non-Markovian nature of channel noise, theoretical studies have used fractal \cite{Liebovitch1991} and chaotic \cite{Liebovitch1987} models of channel gating. In fractal models of channel gating, the open and closed states are represented as a continuum of conformation states, where the current through a single channel is self-similar in different time scales \cite{Liebovitch1987}. In these models, the longer the channel resides in any state, the less likely it is per unit time to exit that state. Alternatively, studies have used chaotic models of channel gating where transitions between kinetic states emerge from deterministic forces instead of random fluctuations of the channel protein \cite{Liebovitch1991}. It has been observed that the Fourier transform of ionic current through BK channels is not a Lorentzian curve as would be expected for statistically independent channels but exhibits a power law with an exponent between $\text{-1}$ and $\text{-2}$ \cite{McGeoch1992}. The $\text{1/f}$ flicker noise can be caused due to various reasons; co-operativity between channels \cite{McGeoch1994}, second-order conformation change in the channel leading to incomplete closure of the pore or obstruction of ion passage across the channel \cite{Mercik2001,Siwy2002}. Whether these observations apply to other classes of voltage-gated ion channels remains unclear \cite{Korn1988}. Because of the limited data from the voltage-gated $Na^+$ and delayed rectifier $K^+$ channels we have modeled, we address only the differences between Langevin and Markov channel noise models. The Langevin and Markov models of channel noise have been used extensively in the literature \cite{Casado2003,Rowat2004,Jo2005,Schmid2004,Schneidman1998,Skaugen1979}, when Markovian and Gaussian statistics are assumed.

Our study indicates that the Langevin model of channel noise is unable to capture the stochastic behaviour of voltage-gated ion channels, although this method has been repeatedly used in the literature \cite{Casado2003,Rowat2004,Jo2005,Schmid2004}. Contrary to popular assumptions \cite{White2000}, the overestimation of information rates by the Langevin model, does not improve in larger area compartments with greater numbers of ion channels; information rate estimates from Markov and Langevin models do not converge even in large compartments. This is true for the largest compartments that we have simulated, which contain 6000 $Na^{+}$ and 1800 $K^{+}$ channels, though it is possible that in even larger compartments the two models, Langevin and Markov, may converge to similar information rate estimates.  

The performance of the Langevin and Markov models has been compared in simulations of other biological systems such as pancreatic $\beta-$cells \cite{Shuai2002,Zhan2007}. These studies have also suggested that the performance of the Langevin model may be inadequate to capture stochastic noise in molecular components of cells. For example, the variance of $[Ca^{2+}]$ flux through $IP_{3}$ receptors in the endoplasmic reticulum of pancreatic $\beta-$cells, calculated using Langevin and Markov models only converged at large numbers of receptors (N=1000) at a variance of zero i.e., the results converged only after the system has become deterministic \cite{Shuai2002}. 
Indeed, even in simulations with 20,000 $IP_{3}$ receptors the output of the Langevin model did not converge with that of the Markov model \cite{Zhan2007}. The underestimation of the channel noise by the Langevin model is a critical issue for studies that investigate noise in molecular systems, especially given the increasing interest in stochastic processes contributing not only to the function of cells and molecular components but also to gene expression. 

\begin{acknowledgments}
We would like to thank Elad Schneidman and Michael Berry for comments during the early stage of the project. This study was supported by the BBSRC (B.S.,S.B.L.), GlaxoSmithKline (B.S.), the Royal Society (J.E.N.), and the Frank Levinson Family Foundation to the STRI Laboratory of Behavior and Evolutionary Neurobiology (J.E.N.).
\end{acknowledgments}

\bibliography{plos_compbiol}

\end{document}